\documentclass[12pt]{article}
\usepackage{graphicx}
\usepackage{epsfig}
\usepackage{epstopdf}
\DeclareGraphicsExtensions{.pdf,.eps,.png,.jpg,.mps}

\def\lsim{\mathrel{\rlap{\lower3pt\hbox{\hskip0pt$\sim$}}
     \raise1pt\hbox{$<$}}}         %less than or approx. symbol
\def\gsim{\mathrel{\rlap{\lower4pt\hbox{\hskip1pt$\sim$}}
     \raise1pt\hbox{$>$}}}         %greater than or approx. symbol

\usepackage{amsmath}
\usepackage{amsfonts}

\begin{document}
\begin{titlepage}

\centerline{\Large \bf Mean-Reversion and Optimization}
\medskip

\centerline{Zura Kakushadze$^\S$$^\dag$$^\ddag$\footnote{\, \tt Email: zura@quantigic.com}}

\bigskip

\centerline{\em $^\S$ Quantigic$^\circledR$ Solutions LLC}
\centerline{\em 1127 High Ridge Road \#135, Stamford, CT 06905\,\,\footnote{\, DISCLAIMER: This address is used by the corresponding author for no
purpose other than to indicate his professional affiliation as is customary in
publications. In particular, the contents of this paper
are not intended as an investment, legal, tax or any other such advice,
and in no way represent views of Quantigic® Solutions LLC,
the website \underline{www.quantigic.com} or any of their other affiliates.
}}
\centerline{\em $^\dag$ Department of Physics, University of Connecticut}
\centerline{\em 1 University Place, Stamford, CT 06901}
\centerline{\em $^\ddag$ Free University of Tbilisi, Business School \& School of Physics}
\centerline{\em 240, David Agmashenebeli Alley, Tbilisi, 0159, Georgia}
\medskip
\centerline{(August 9, 2014; revised September 22, 2014)}

\bigskip
\medskip

\begin{abstract}
{}The purpose of these notes is to provide a systematic quantitative framework -- in what is intended to be a ``pedagogical" fashion -- for discussing mean-reversion and optimization. We start with pair trading and add complexity by following the sequence ``mean-reversion via demeaning $\rightarrow$ regression $\rightarrow$ weighted regression $\rightarrow$ (constrained) optimization $\rightarrow$ factor models". We discuss in detail how to do mean-reversion based on this approach, including common pitfalls encountered in practical applications, such as the difference between maximizing the Sharpe ratio and minimizing an objective function when trading costs are included. We also discuss explicit algorithms for optimization with linear costs, constraints and bounds. We also illustrate our discussion on an explicit intraday mean-reversion alpha.
\end{abstract}
\medskip
\end{titlepage}

\newpage

\section{Introduction and Summary}

{}Statistical Arbitrage (StatArb) ``refers to highly technical short-term mean-reversion strategies involving large numbers of securities (hundreds to thousands, depending on the amount of risk capital), very short holding periods (measured in days to seconds), and substantial computational, trading, and information technology (IT) infrastructure" (Lo, 2010). So, what is this ``mean-reversion"?\footnote{\, ``Mean-reversion strategy" is mostly trader lingo -- which is what the author is accustomed to. Academic finance literature mostly uses ``contrarian investment strategy" instead. This paper uses the term ``mean-reversion (strategy)" throughout.} The basic idea is simple: some quantities are historically correlated, sometimes these correlations are temporarily undone by some unusual market conditions, but one expects -- or rather hopes -- that the correlation will be restored in the future. StatArb tries to capture a profit from such temporary mispricings.

{}The purpose of these notes is to provide a systematic quantitative framework -- in what is intended to be a ``pedagogical" fashion -- for discussing mean-reversion and optimization. There are a myriad ways of doing ({\em i.e.}, implementing) mean-reversion. One such approach can be schematically described via a sequence ``mean-reversion via demeaning $\rightarrow$ regression $\rightarrow$ weighted regression $\rightarrow$ (constrained) optimization $\rightarrow$ factor models". These notes follow precisely this sequence, starting from the most basic form of StatArb, pair trading, and gradually adding complexity. This naturally introduces mean-reversion around means of returns, regression and, ultimately, optimization and factor models -- via the observation that weighted regression is nothing but a zero specific risk\footnote{\, ``Specific risk" is multi-factor risk model terminology. Some may prefer ``idiosyncratic risk".} limit of optimization with a factor model.

{}Within this framework we discuss various important intricacies and pitfalls that arise in practical applications, often overlooked, deemphasized and/or not addressed, including in commercially available offerings. Should regression weights used in conjunction with optimization be based on historical or specific risk? How should one optimize regressed returns? How does one include constraints into optimization? Is optimization based on objective function minimization the same as maximizing the Sharpe ratio once costs are included? How does one optimize with linear costs, constraints and bounds? {\em Etc.} These are some of the topics we discuss in these notes -- systematically and ``pedagogically", we hope.

{}The organization of these notes is as follows. Section 2 discusses mean reversion: pair trading $\rightarrow$ multiple stocks $\rightarrow$ multiple binary clusters (industries) $\rightarrow$ regression $\rightarrow$ non-binary generalization $\rightarrow$ weighted regression. Section 3 discusses optimization: maximizing Sharpe ratio $\rightarrow$ adding multiple linear constraints (including dollar neutrality) $\rightarrow$ regression as a limit of optimization $\rightarrow$ factor models $\rightarrow$ optimization with a factor model with linear constraints (including pitfalls). Section 4 is an ``intermezzo" of sorts, which is kept on the lighter side, to help digest Sections 2 and 3 before adding even more complexity in Sections 5 and 6. Section 5 discusses optimization with constraints and costs, including the difference between Sharpe ratio maximization and minimizing an objective function, and when the latter can be used as an approximation for the former. Section 6 discusses explicit algorithms for optimization with linear costs, constraints and bounds, including in the context of factor models. Section 7 illustrates the regression approach discussed in Section 2 by giving an explicit example of an intraday mean-reversion alpha (with a 5-year simulated performance), based on overnight returns and an industry classification, together with additional bells and whistles for risk management and dealing with outliers. Section 8 contains brief concluding remarks.

\section{Mean-Reversion}\label{sec.mean.rev}

\subsection{Pair Trading}

{}Often, when explaining StatArb popularly, a pair trading example is given. In a nutshell, it goes as follows. Suppose you have two historically correlated stocks in the same sector, stock A and stock B ({\em e.g.}, Exxon Mobil (XOM) and Royal Dutch Shell (RDS.A)). If, temporarily, stock A moves up (A is rich) while stock B moves down (B is cheap), the pair trading strategy amounts to shorting A and buying B in such proportion that the total position is dollar neutral. Dollar neutrality ensures that the position is (approximately) insensitive to overall market movements -- it is simply a hedge against market risk.

{}Intuitively, this all makes sense assuming the spread between A and B converges back to its historical values. This is the essence of mean-reversion. The money is made from a temporary mispricing in A and B. However, upon a second look, one may ask: How do I know if A is rich and B is cheap? Indeed, first, A and B typically trade at different prices to begin with. Second, the prices of A and B are not constant on average -- typically, albeit not always, they each have an upward drift. So, how does one quantify ``rich" and "cheap" in pair trading?

\subsection{Returns, Not Prices}

{}It is not prices but returns that define ``rich" and ``cheap". The idea is that on average stocks A and B are expected to move in sync. Say both move up. If A moves up more than B on a relative basis to their respective prices, then A is rich and B is cheap. Let $P_A(t_1)$ and $P_B(t_1)$ be the prices of A and B at time $t_1$, and let $P_A(t_2)$ and $P_B(t_2)$ be the prices of A and B at a later time $t_2$. ({\em E.g.}, $t_1$ can be yesterday's close -- with $P_A(t_1)$ and $P_B(t_1)$ adjusted for any splits and dividends if the ex-date is today -- and $t_2$ can be today's open.) The corresponding returns are
\begin{eqnarray}
 &&R_A = {P_A(t_2)\over P_A(t_1)} - 1\\
 &&R_B = {P_B(t_2)\over P_B(t_1)} - 1
\end{eqnarray}
Since typically these returns are small, we can use an alternative definition:
\begin{eqnarray}
 &&R_A \equiv \ln\left({P_A(t_2)\over P_A(t_1)}\right)\\
 &&R_B \equiv \ln\left({P_B(t_2)\over P_B(t_1)}\right)
\end{eqnarray}
So, the mean-reversion idea in pair trading can now be quantified as follows. If $R_A > R_B$, then A is rich, B is cheap, short A and buy B.

{}We can conveniently restate this using the demeaned returns ${\widetilde R}_A$ and ${\widetilde R}_B$:
\begin{eqnarray}
 &&{\overline R} \equiv {1\over 2}\left(R_A + R_B\right)\\
 &&{\widetilde R}_A \equiv R_A - {\overline R}\\
 &&{\widetilde R}_B \equiv R_B - {\overline R}
\end{eqnarray}
where ${\overline R}$ is the mean return.\footnote{\, Here and in the following ${\overline R}$ refers to the {\em cross-sectional} mean return (not the time series mean return). Also, ${\widetilde R}_A$, ${\widetilde R}_B$ and ${\widetilde R}_i$ (see below) refer to the deviation from the mean return ${\overline R}$.} Now a stock is rich if its demeaned return is positive, and it is cheap if its demeaned return is negative. So, assuming the returns have been demeaned, we short positive return stocks and buy negative return stocks.

{}In the case of 2 stocks, the numbers of shares $Q_i$, $i=A,B$ to short/buy are fixed by the total desired dollar investment $I$ and the requirement of dollar neutrality:
\begin{eqnarray}
 &&P_A~\left|Q_A\right| + P_B~\left|Q_B\right| = I\\
 &&P_A~Q_A + P_B~Q_B = 0
\end{eqnarray}
where $P_i$ are the prices at the time the position is established, $Q_i < 0$ for short-sales, and $Q_i > 0$ for buys. Here we assume no leverage and 0 margins (see footnote \ref{foot.margins}).

\subsection{Generalization to Multiple Stocks}\label{sub1.3}

{}What if we have more than two historically correlated stocks in the same sector? ({\em e.g.}, Exxon Mobil, Royal Dutch Shell, Total (TOT), Chevron (CVX) and BP (BP)). While we can do pair trading for each pair of stocks from such a set, can we have a mean-reversion strategy for the entire set? Demeaned returns make this a breeze.

{}Let $R_i$, $i=1,\dots,N$ be the returns for our historically correlated $N$ stocks:
\begin{eqnarray}
 &&R_i = \ln\left({P_i(t_2)\over P_i(t_1)}\right)\\
 &&{\overline R} \equiv {1\over N}\sum_{i=1}^N R_i\\
 &&{\widetilde R}_i \equiv R_i - {\overline R}
\end{eqnarray}
So, following our intuition from the 2-stock example, we can short stocks with positive ${\widetilde R}_i$ and buy stocks with negative ${\widetilde R}_i$. We have the following conditions:\footnote{\label{foot.margins}\, We assume no leverage and 0 margins. Nontrivial leverage simply rescales the investment level $I$. If margins are present, on top of $I$ invested in stocks, we need an additional amount $I^\prime$ to maintain margins, which simply reduces the strategy return due to the borrowing interest rate.}
\begin{eqnarray}\label{tot.inv}
 &&\sum_{i=1}^N P_i~\left|Q_i\right| = I\\
 &&\sum_{i=1}^N P_i~Q_i = 0\label{dollar.neutral}
\end{eqnarray}
Here: $I$ is the total desired dollar investment; (\ref{dollar.neutral}) is dollar neutrality; $Q_i<0$ for short-sales; $Q_i>0$ for buys; $P_i$ are the prices at the time the position is established. We have 2 equations and $N > 2$ unknowns. So, we need to specify how to fix $Q_i$.

{}A simple way of specifying $Q_i$ is to have the dollar positions
\begin{equation}
 D_i\equiv P_i~Q_i
\end{equation}
proportional to the demeaned returns:
\begin{equation}\label{D.i}
 D_i = -\gamma~{\widetilde R}_i
\end{equation}
where $\gamma > 0$ (recall that we short ${\widetilde R}_i>0$ stocks and buy ${\widetilde R}_i<0$ stocks). Then (\ref{dollar.neutral}) is automatically satisfied as $\sum_{i=1}^N {\widetilde R}_i = 0$ by definition, while  (\ref{tot.inv}) fixes $\gamma$:
\begin{equation}
 \gamma = {I \over \sum_{i=1}^N \left|{\widetilde R}_i\right|}
\end{equation}
Eq. (\ref{D.i}) defines {\em one} mean-reversion strategy. There are a myriad of them. One drawback of (\ref{D.i}) is that, by construction, on average it will take larger positions in more volatile stocks (as volatile stocks on average have larger $|{\widetilde R}_i|$). Below we will discuss risk management and other ways of constructing $D_i$, {\em i.e.}, other mean-reversion strategies. Let us discuss a further generalization first.

\subsection{Generalization to Multiple Clusters}\label{sub1.4}

{}We will refer to each group of stocks for which we can perform the analysis of the previous subsection as ``clusters". Depending on a given industry classification scheme, such clusters are called different names, such as industries, sub-industries, {\em etc.}\footnote{\, {\em E.g.}, we could have one group of stocks from the oil sector, the second group from technology, and the third group from, say, healthcare.} Let there be $K$ clusters labeled by $A=1,\dots,K$. Let $\Lambda_{iA}$ be an $N\times K$ matrix such that if the stock labeled by $i$ ($i=1,\dots,N$) belongs to the cluster labeled by $A$, then $\Lambda_{iA}=1$; otherwise, $\Lambda_{iA}=0$. We will assume that each and every stock belongs to one and only one cluster (so there are no empty clusters), {\em i.e.},
\begin{eqnarray}
 &&N_A \equiv \sum_{i=1}^N \Lambda_{iA} > 0\\
 &&N = \sum_{A=1}^K N_A
\end{eqnarray}
We have
\begin{eqnarray}
 &&\Lambda_{iA} = \delta_{G(i), A}\\
 &&G:\{1,\dots, N\} \mapsto \{1,\dots, K\}
\end{eqnarray}
where $G$ is the map between stocks and clusters. The matrix $\Lambda_{iA}$ is referred to as the loadings matrix. The Kronecker delta $\delta_{ab} = 1$ if $a=b$, and $\delta_{ab} = 0$ if $a\neq b$.

{}Mean-reversion can be done separately for each cluster as the clusters do not overlap. However, for further generalization, it is convenient to write the demeaned returns in a compact form, for all clusters at once. This brings in regression.

\subsection{Regression}
{}Let $R_i$ be the stock returns. Consider a linear regression of $R_i$ over $\Lambda_{iA}$ (without intercept and with unit weights -- see below). In R notation:\footnote{\, The R Package for Statistical Computing. Also, ``$\sim$" in (\ref{R.lm}) is R notation for a linear model.}
\begin{equation}\label{R.lm}
 R \sim -1 + \Lambda
\end{equation}
where, in matrix notation, $R$ is the $N$-vector $R_i$, and $\Lambda$ is the $N\times K$ loadings matrix $\Lambda_{iA}$. Explicitly, we have
\begin{equation}\label{R.Lambda}
 R_i = \sum_{A=1}^K \Lambda_{iA}~f_A + \varepsilon_i
\end{equation}
where $f_A$ are the regression coefficients given by (in matrix notation)
\begin{eqnarray}
 &&f = Q^{-1}~\Lambda^T~R\\
 &&Q \equiv \Lambda^T~\Lambda
\end{eqnarray}
and $\varepsilon_i$ are the regression residuals. In the case of binary $\Lambda_{iA}$ we introduced in the previous subsection, these residuals are nothing but the returns $R_i$ demeaned w.r.t. to the corresponding cluster:
\begin{eqnarray}
 &&\varepsilon = R - \Lambda~Q^{-1}~\Lambda^T~R\\
 &&Q_{AB} = N_A~\delta_{AB}\\
 &&{\overline R}_A\equiv {1\over N_A}\sum_{j\in J_A} R_j\\
 &&\varepsilon_i = R_i - {\overline R}_{G(i)} = {\widetilde R}_i
\end{eqnarray}
where ${\overline R}_A$ is the mean return for the cluster labeled by $A$, and ${\widetilde R}_i$ is the demeaned return obtained by subtracting from $R_i$ the mean return for the cluster labeled by $A=G(i)$ to which the stock labeled by $i$ belongs: $G(i)=A: i\in J_A\subset \{1,\dots,N\}$.

{}So, the demeaned returns ${\widetilde R}_i$ are given by the residuals of a regression (without intercept and with unit weights) of the returns $R_i$ over the loadings matrix $\Lambda_{iA}$. This result allows to further generalize the above construction. But first some additional observations are in order.

{}Note that
\begin{equation}\label{ind.neutral}
 \sum_{i=1}^N {\widetilde R}_i~\Lambda_{iA} = 0,~~~A=1,\dots,K
\end{equation}
{\em I.e.}, the demeaned returns are cluster neutral, or, if the clusters are referred to as industries, they are industry neutral. In this case this is simply the statement that for each cluster the sum of the demeaned returns over all stocks in such cluster vanishes, which follows from the fact that the returns are demeaned w.r.t. each cluster. However, in a more general case (see below) this is a more nontrivial condition.

{}Also, note that we automatically have
\begin{equation}\label{d.n.1}
 \sum_{i=1}^N {\widetilde R}_i~\nu_i = 0
\end{equation}
where $\nu_i\equiv 1$, $i=1,\dots, N$, {\em i.e.}, the $N$-vector $\nu$ is the unit vector. In the regression language, $\nu$ is referred to as the intercept. Above we did not have to add the intercept to the loadings matrix because it is already subsumed in it:
\begin{equation}
 \sum_{A=1}^K \Lambda_{iA} = \nu_i
\end{equation}
However, in the general case, to have (\ref{d.n.1}), we would need to add the intercept as a column in the loadings matrix (see below). Recall that (\ref{d.n.1}) is the same as dollar neutrality in the strategy (\ref{D.i}), but generally dollar neutrality does not require (\ref{d.n.1}).

\subsection{Non-binary Generalization}

{}The conditions (\ref{ind.neutral}) satisfied by the demeaned returns in the binary loadings matrix case simply mean that these returns are cluster neutral, {\em i.e.}, orthogonal to the corresponding $N$-vectors $v^{(A)}$, where $v^{(A)}_i \equiv \Lambda_{iA}$. That is, in matrix notation
\begin{equation}
 {\widetilde R}^T~v^{(A)} = 0
\end{equation}
This orthogonality can be defined for any loadings matrix, not just a binary one.

{}This leads us to a generalization where the loadings matrix, call it $\Omega_{iA}$, may have some binary columns, but generally it need not. The binary columns, if any, are interpreted as industry (cluster) based risk factors; the non-binary columns are interpreted as some non-industry based risk factors; and the orthogonality condition
\begin{equation}
 \sum_{i=1}^N {\widetilde R}_i~\Omega_{iA},~~~A=1,\dots,K
\end{equation}
is simply the requirement that the twiddled returns ${\widetilde R}_i$ -- which we will no longer refer to as ``demeaned returns" (for the loadings matrix is no longer necessarily binary), but instead we will refer to them as ``regressed returns" -- are the residuals of the regression (without intercept and with unit weights) of $R_i$ over $\Omega_{iA}$:
\begin{eqnarray}\label{R.tilde}
 &&{\widetilde R} \equiv R - \Omega~Q^{-1}~\Omega^T~R\\
 &&Q \equiv \Omega^T~\Omega
\end{eqnarray}
Note that we no longer necessarily have the property (\ref{d.n.1}). If this property is desired, it can be achieved by including the intercept in the regression. {\em I.e.}, in the R notation the regression (now with intercept, but still with unit weights) is
\begin{equation}
 R \sim \Omega
\end{equation}
In terms of (\ref{R.tilde}) this simply amounts to adding a unit column to $\Omega$, so now we have $\Omega_{iA_1} \equiv \nu_i = 1$, $i=1,\dots,N$ for some column labeled by $A_1$.

\subsection{Weighted Regression}\label{sub1.7}

{}In Subsection \ref{sub1.3} we discussed a simple strategy (\ref{D.i}), where the desired dollar holdings $D_i$ are proportional to ${\widetilde R}_i$. One potential ``shortcoming" in this strategy is that on average its positions will be dominated by volatile stocks. One idea for reducing this exposure to volatility is to divide ${\widetilde R}_i$ by $\sigma_i$ or $\sigma_i^2$ (or some other power of $\sigma_i$), where $\sigma_i$ is, {\em e.g.}, the historical volatility of ${\widetilde R}_i$ or, more simply,\footnote{\, We will discuss these subtleties below, when we discuss factor models and optimization.} of $R_i$, {\em i.e.}, variances $\sigma_i^2 \equiv C_{ii}$ are the diagonal elements of the sample covariance matrix
\begin{equation}
 C_{ij} \equiv\langle R_i, R_j\rangle
\end{equation}
where the covariances $\langle\cdot,\cdot\rangle$ are computed over the corresponding time series of $R_i$.

{}Here a few remarks are in order. First, if we take, say, ${\widehat R}_i \equiv {\widetilde R}_i / \sigma_i^2$, even if ${\widetilde R}^T~\nu = 0$, generally we do not have ${\widehat R}^T~\nu = 0$, so using ${\widehat R}_i$ instead of ${\widetilde R}_i$ in (\ref{D.i}) would generally spoil the dollar neutrality property ({\em i.e.}, that $\sum_{i=1}^N D_i = 0$). We need to deal with this somehow. Second, should we take ${\widehat R}_i \equiv {\widetilde R}_i /\sigma_i$ or ${\widehat R}_i \equiv {\widetilde R}_i /\sigma_i^2$ or something else? The preferred answer to the last question is that we need to take ${\widehat R}_i \equiv {\widetilde R}_i /\sigma_i^2$ -- or, more precisely, its variant we will come to in a moment -- and the reason for this will become clear when we discuss optimization. This may appear a bit odd at first as the return with the risk scaled out of it should be ${\widetilde R}_i /\sigma_i$, not ${\widetilde R}_i /\sigma_i^2$. However, the extra suppression by another factor of $\sigma_i$ is what maximizes the Sharpe ratio, which we will discuss in more detail below. For now, we will take this for granted and suppress the return ${\widetilde R}_i$ by $\sigma_i^2$. All we have to figure out is how to make sure that we do not spoil dollar neutrality in the process.

{}One answer is given by weighted regression, where $R$ is regressed over $\Omega$ with weights $z_i$. We have
\begin{eqnarray}\label{w.reg.1}
 &&\varepsilon \equiv R - \Omega~Q^{-1}~\Omega^T~Z~R\\
 \label{w.reg.2}
 &&Z \equiv\mbox{diag}(z_i)\\
 \label{w.reg.3}
 &&Q\equiv \Omega^T~Z~\Omega\\
 \label{w.reg.4}
 &&{\widetilde R} \equiv Z~\varepsilon
\end{eqnarray}
Here $\varepsilon_i$ are the residuals of the weighted regression. Also, note that
\begin{equation}
 \sum_{i=1}^N{\widetilde R}_i~\Omega_{iA} = 0,~~~A=1,\dots,K
\end{equation}
If the intercept is included in $\Omega_{iA}$, then we automatically have $\sum_{i=1}^N {\widetilde R}_i = 0$. Also, if we take $z_i = 1/\sigma_i^2$, then ${\widetilde R}_i$ will be suppressed by $\sigma_i^2$ compared with the case of the regression with unit weights. So, now our simple strategy (\ref{D.i}) is not only dollar neutral but has risk management built into it. The resulting holdings are neutral w.r.t. the risk factors described by the loadings matrix $\Omega_{iA}$, and furthermore are no longer dominated by volatile stock holdings. This is now a real mean-reversion strategy. We will take it a step further in the next section.

\subsection{Remarks}

{}As mentioned above, (\ref{D.i}) is only one of a myriad ways of specifying $D_i$ given the regressed returns ${\widetilde R}_i$. If the regression includes the intercept, ${\widetilde R}_i$ have 0 cross-sectional mean, so the strategy defined by (\ref{D.i}) is automatically dollar neutral. Also, if the regression is weighted as in Subsection \ref{sub1.7}, contributions from high volatility stocks are weighted down thereby providing risk management.\footnote{\, As we discuss in the next section, this case is a certain limit of optimization.} Here, for illustrative purposes only (and not as an exhaustive survey), we discuss some other mean-reversion strategies, {\em i.e.}, other ways of specifying $D_i$.

{}One simple example is to have ``equally weighted" $D_i$
\begin{equation}\label{D.i.sign}
 D_i = -\gamma~\mbox{sign}({\widetilde R}_i)
\end{equation}
where $\gamma >0$, {\em i.e.}, we buy stocks with negative regressed returns and sell stocks with positive regressed returns, all with the same absolute dollar amount equal $\gamma = I/N$ (this equality follows from (\ref{tot.inv})). This strategy has some evident ``shortcomings". First, it is not necessarily dollar neutral:
\begin{equation}\label{mishedge}
 \sum_{i=1}^N D_i = {{N_- - N_+}\over N}~I
\end{equation}
where $N_+$ is the number of stocks with positive regressed returns and $N_-$ is the number of stocks with negative regressed returns, and generally $N_+ \neq N_-$. If $N$ is large, then {\em assuming} a normal distribution for $N_- - N_+$ with mean 0 and standard deviation of order $\sqrt{N}$, the mishedge (\ref{mishedge}) is of order $I/\sqrt{N}$. For $N\sim 2,500$, this is of order 2\%, which may be unacceptably large. To achieve dollar neutrality, we can modify the values of some $D_i$, {\em e.g.}, by setting some of them to zero. One then needs to decide which values to set to zero. This brings us to the second ``shortcoming" in this strategy: $\mbox{sign}(x)$ is discontinuous across $x = 0$, so for small ${\widetilde R}_i$ the sign of $D_i$ can flip even with small fluctuations.\footnote{\, Be it due to changes from one day to another, or due to computational uncertainties, {\em etc.}} This instability can result in unnecessary portfolio turnover (overtrading) and additional trading costs, and generally diminish the performance of the strategy. One way to ``smooth" this out is to approximate $\mbox{sign}(x)$ via, {\em e.g.}, $\tanh(x/\kappa)$:
\begin{equation}\label{D.i.tanh}
 D_i = -\gamma~\tanh({\widetilde R}_i / \kappa)
\end{equation}
where $\kappa$ is the {\em cross-sectional} standard deviation of ${\widetilde R}_i$. Then, for $|{\widetilde R}_i| \ll \kappa$ this approximately reduces to (\ref{D.i}), whereas for $|{\widetilde R}_i| \gsim \kappa$ the dollar holdings are ``squashed". If the regression has unit weights, then on average $|{\widetilde R}_i|$ are larger for more volatile stocks compared with less volatile stocks, and using (\ref{D.i.tanh}) amounts to suppressing the contributions from less volatile stocks while ``equally" weighting the contributions from more volatile stocks. As mentioned above, this may not be desirable from the risk management viewpoint. If the regression is weighted with $z_i = 1/\sigma_i^2$, then on average $|{\widetilde R}_i|$ are suppressed for more volatile stocks compared with less volatile stocks, so using (\ref{D.i.tanh}) amounts to suppressing the contributions from more volatile stocks, while ``equally" weighting the contributions from less volatile stocks. In this case we can achieve dollar neutrality by setting to zero and/or appropriately scaling down the absolute values of $D_i$ for more volatile stocks. Let us mention that (\ref{D.i.tanh}) generally is farther away than (\ref{D.i}) (assuming ${\widetilde R}_i$ are based on a weighted regression with $z_i=1/\sigma_i^2$) from the optimized solution we discuss in the next section.

{}If one contemplates (\ref{D.i.sign}) (or (\ref{D.i.tanh}) as its ``smoothed out" version), one may also explore the opposite direction and consider, {\em e.g.},
\begin{equation}\label{D.i.RR}
 D_i = - \gamma~{\widetilde R}_i~|{\widetilde R}_i|
\end{equation}
Here ${\widetilde R}_i$ are based on a weighted regression -- otherwise the portfolio would be too volatile. In fact, more generally one can consider strategies with
\begin{equation}\label{D.i.Rf}
 D_i = - \gamma~{\widetilde R}_i~f({\widetilde R}_i)
\end{equation}
where $f(x)$ is some function. Such ``nonlinear alphas" are commonly used in quant trading. Note that, as for (\ref{D.i.sign}) and (\ref{D.i.tanh}), (\ref{D.i.RR}) and more generally (\ref{D.i.Rf}) require additional ``gymnastics" to achieve dollar neutrality. There is no ``magic prescription" for picking $f(x)$ in (\ref{D.i.Rf}). In practice at any given time one picks alphas that backtest well, and alphas are ephemeral by nature -- alphas that work now may not work 6 months from now. This is an ever-changing empirical game, not a theoretical one.

{}This brings us to yet another commonly used way of specifying $D_i$: ranking. Instead of using continuous functions such as, {\em e.g.}, (\ref{D.i}) or (\ref{D.i.Rf}), one can, {\em e.g.}, rank stocks cross-sectionally by $|{\widetilde R}_i|$. Let this integer rank be $r_i$. Then we can take, {\em e.g.}:
\begin{equation}
 D_i = -\gamma~\mbox{sign}({\widetilde R}_i)~r_i
\end{equation}
Alternatively, we can set $D_i$ to 0 for the stocks with $r_i < r_*$. Various comments we made above relating to risk management and dollar neutrality also apply to alphas based on ranking. Furthermore, one can consider nonlinear functions of $r_i$.

{}In this regard, let us also mention that above we treat dollar neutrality symmetrically between long and short holdings. There are other possibilities here too. {\em E.g.}, we can go long cash ({\em i.e.}, stocks -- more trader lingo) with $D_i$ specified via, say, (\ref{D.i}), and short the same dollar amount of futures for some diversified index, {\em e.g.} S\&P500 -- this would be a so-called S\&P outperformance portfolio. In this case we have lower bounds $D_i\geq 0$. Similarly, instead of shorting futures, we could short a tracking portfolio for the index, {\em e.g.}, a minimum variance portfolio, whose weights are independent of the stock expected returns.\footnote{\, In this case, the actual portfolio consists of net long or short positions for individual stocks arising from long positions $D_i$ and short positions from the minimum variance portfolio.} Instead of limiting the short position to a tracking portfolio for an index, one can consider a minimum variance portfolio for some (diversified) proprietary trading universe. As mentioned above, there are many ways of doing mean-reversion. Here we focus on the sequence ``mean-reversion via demeaning $\rightarrow$ regression $\rightarrow$ weighted regression $\rightarrow$ (constrained) optimization $\rightarrow$ factor models", which brings us to our next topic -- optimization.

\section{Optimization}

\subsection{Maximizing Sharpe Ratio}

{}Let $C_{ij}$ be the sample covariance matrix of the $N$ time series of stock returns $R_i(t_s)$, $s=0,1,\dots,M$, where $t_0$ is the most recent time. Below $R_i$ refers to $R_i(t_0)$. Let $\Psi_{ij}$ be the corresponding correlation matrix, {\em i.e.},
\begin{equation}
 C_{ij} = \sigma_i~\sigma_j~\Psi_{ij}
\end{equation}
where $\Psi_{ii} = 1$. For the sake of definiteness, let us assume that $R_i$ are daily returns, albeit this is not a critical assumption.

{}As above, let $D_i$ be the dollar holdings in our portfolio. The portfolio P\&L, volatility and Sharpe ratio are given by
\begin{eqnarray}
 &&P = \sum_{i=1}^N R_i~D_i\\
 &&V = \sqrt{\sum_{i,j=1}^N C_{ij}~D_i~D_j}\\
 &&S = {P \over V}
\end{eqnarray}
Instead of dollar holdings $D_i$, it is more convenient to work with dimensionless holding weights (not to be confused with the regression weights $z_i$)
\begin{equation}
 w_i \equiv {D_i\over I}
\end{equation}
where $I$ is the investment level. The holding weights satisfy the condition
\begin{equation}\label{w.norm}
 \sum_{i=1}^N \left|w_i\right| = 1
\end{equation}
They are positive for long holdings and negative for short holdings.

{}In terms of the holding weights, the P\&L and volatility are given by
\begin{eqnarray}
 &&P = I~{\widetilde P} \equiv I~\sum_{i=1}^N R_i~w_i\\
 &&V = I~{\widetilde V} \equiv I~\sqrt{\sum_{i,j=1}^N C_{ij}~w_i~w_j}
\end{eqnarray}
To determine the weights, often one requires that the Sharpe ratio be maximized:
\begin{equation}\label{max.sharpe}
 S \rightarrow \mbox{max}
\end{equation}
Assuming (for now) that there are no additional conditions on $w_i$ ({\em e.g.}, upper or lower bounds), the solution to (\ref{max.sharpe}) in the absence of costs is given by
\begin{equation}\label{w.max.sharpe}
 w_i = \gamma \sum_{j=1}^N C^{-1}_{ij}R_j
\end{equation}
where $C^{-1}$ is the inverse of $C$, and the normalization coefficient $\gamma$ is determined from (\ref{w.norm}). Invertibility of $C$ should not be taken for granted and we will discuss this issue a bit later. However, for now, let us assume that $C$ is invertible.

{}One immediate consequence of (\ref{w.max.sharpe}) is that these holding weights generically do not correspond to a dollar neutral portfolio. {\em E.g.}, if $C_{ij}$ is diagonal and all $R_i > 0$, then all $w_i > 0$. More generally, there is no reason why $\sum_{i=1}^N w_i$ should vanish. So, if we wish to have a dollar neutral portfolio, we need to maximize the Sharpe ratio subject to the dollar neutrality constraint.

\subsection{Linear Constraints; Dollar Neutrality}

{}Dollar neutrality can be achieved as follows. First, note that the Sharpe ratio is invariant under the simultaneous rescalings of all holding weights $w_i\rightarrow \zeta~w_i$, where $\zeta >0$. Because of this scale invariance, the Sharpe ratio maximization problem can be recast in terms of minimizing a quadratic objective function:
\begin{eqnarray}
 &&g(w, \lambda) \equiv {\lambda\over 2} \sum_{i,j=1}^N C_{ij}~w_i~ w_j - \sum_{i=1}^N R_i~w_i\\
 &&g(w, \lambda)\rightarrow\mbox{min}
\end{eqnarray}
where $\lambda > 0$ is a parameter, and minimization is w.r.t. $w_i$. The solution is given by
\begin{equation}
 w_i = {1\over\lambda}~\sum_{j=1}^N C^{-1}_{ij}~R_j
\end{equation}
and $\lambda$ is fixed via (\ref{w.norm}). The objective function approach is convenient if we wish to impose constraints on $w_i$, {\em e.g.}, the dollar neutrality constraint. We introduce an $N\times m$ matrix $Y_{ia}$ and $m$ Lagrange multipliers $\mu_a$, $a=1,\dots, m$:
\begin{eqnarray}
 &&g(w, \mu, \lambda) \equiv {\lambda\over 2} \sum_{i,j=1}^N C_{ij}~w_i~ w_j - \sum_{i=1}^N R_i~w_i - \sum_{a=1}^m\sum_{i=1}^N w_i~Y_{ia}~\mu_a\\
 &&g(w, \mu, \lambda)\rightarrow\mbox{min}
\end{eqnarray}
Minimization w.r.t. $w_i$ and $\mu_a$ now gives the following equations:
\begin{eqnarray}\label{opt.w.7}
 &&\lambda~\sum_{j=1}^N C_{ij}~w_j = R_i + \sum_{a=1}^m Y_{ia}~\mu_a\\
 &&\sum_{i=1}^N w_i~Y_{ia} = 0\label{opt.mu.7}
\end{eqnarray}
So, we have $m$ homogeneous linear constraints (\ref{opt.mu.7}). If $Y_{ia_1} \equiv \nu_i = 1$, $i=1,\dots,N$ for some $a_1\in\{1,\dots,m\}$, then we have dollar neutrality. Note that $m$ can be 1.

{}The solution to (\ref{opt.w.7}) and (\ref{opt.mu.7}) is given by (in matrix notation):
\begin{eqnarray}
 &&w = {1\over\lambda}~\left[C^{-1} - C^{-1}~Y~(Y^T~C^{-1}~Y)^{-1}~Y^T~C^{-1}\right]~R
 \label{w.opt.const}\\
 &&\mu = - (Y^T~C^{-1}~Y)^{-1}~Y^T~C^{-1}~R
 \label{mu.opt.const}
\end{eqnarray}
As before, $\lambda$ is fixed via (\ref{w.norm}). The solution (\ref{w.opt.const}) and (\ref{mu.opt.const}) can be rewritten as follows:
\begin{eqnarray}
 &&\omega = {1\over\lambda}~\Gamma^{-1}~\rho\\
 &&\omega^T \equiv \left(w^T,~ -\lambda^{-1}\mu^T\right)\\
 &&\rho^T \equiv \left(R^T,~O^T\right)\\
 &&\Gamma \equiv\left( \begin{array}{ll}
               C & Y\\
              Y^T & \mathbb{O} \\
               \end{array}\right)
\end{eqnarray}
{\em I.e.}, $\omega$ and $\rho$ are $(N+m)$-vectors, and $\Gamma$ is an $(N+m)\times(N+m)$ matrix; $O$ is a nil $m$-vector, and $\mathbb{O}$ is a nil $m\times m$ matrix. Thus, linear constraints can be dealt with by simply enlarging the covariance matrix as above.\footnote{\, Above we considered homogeneous constraints (\ref{opt.mu.7}). Technically, the same trick can be applied to inhomogeneous constraints of the form $\sum_{i=1}^N w_i~Y_{ia} + y_a = 0$. Everything goes through as above, except that now we have $\rho_a = -\lambda~y_a$. However, while this will give the correct solution to the minimization of the objective function, this is no longer necessarily the same as maximizing the Sharpe ratio with constraints: the latter explicitly break the invariance under the rescalings $w_i\rightarrow \zeta~w_i$ (unless $y_a\equiv 0$), which is what allowed us to rewrite the Sharpe ratio maximization problem in terms of the objective function minimization problem, whereby $\lambda$ is fixed via (\ref{w.norm}). In the presence of inhomogeneous constraints this is no longer the case and some additional care is needed -- see Section \ref{sec.opt.cost}. We will not need inhomogeneous constraints here, however.}

\subsection{Regression as Constrained Diagonal Optimization}

{}Let us now consider the case where the covariance matrix is diagonal: $C_{ij}= \sigma^2_i~\delta_{ij}$. Then (\ref{w.opt.const}) reads
\begin{equation}
 w = {1\over\lambda}~\left[Z - Z~Y~(Y^T~Z~Y)^{-1}~Y^T~Z\right]~R = {1\over\lambda}~Z~\varepsilon = {1\over\lambda}~{\widetilde R}
\end{equation}
Here $Z \equiv \mbox{diag}(1/\sigma_i^2)$, $\varepsilon_i$ are the residuals of the weighted regression with weights $z_i=1/\sigma^2_i$ of $R_i$ over the $N\times m$ matrix $Y_{ia}$ (without intercept -- unless the intercept is already included in $Y_{ia}$, that is). This is the same weighted regression we discussed in Subsection \ref{sub1.7}. So, diagonal (meaning, with diagonal covariance matrix) constrained optimization is the same as the weighted regression with the loadings matrix $\Omega$ identified with the constraint matrix $Y$, and the regression weights $z_i$ (not to be confused with the holding weights $w_i$) identified with inverse variances of the returns $R$. Furthermore, the holding weights $w_i$ are given by the regressed returns ${\widetilde R}_i$ up to a normalization factor fixed via (\ref{w.norm}). If the constraint matrix contains the intercept (the unit vector), then the holding weights correspond to a dollar neutral portfolio.

\subsection{Regression as Limit of Optimization}\label{sub2.5}

{}Weighted regression (\ref{w.reg.1}), (\ref{w.reg.2}), (\ref{w.reg.3}) and (\ref{w.reg.4}) has a structure such that it is actually related to factor models. Consider an auxiliary matrix
\begin{eqnarray}
 && \Theta \equiv \Xi + \zeta~\Omega~\Omega^T\\
 && \Xi \equiv Z^{-1}
\end{eqnarray}
where $\zeta$ is a parameter. The inverse reads:
\begin{eqnarray}
 &&\Theta^{-1} = Z - \zeta~Z~\Omega~{\widetilde Q}^{-1}~\Omega^T~Z\\
 &&{\widetilde Q}_{AB} \equiv \delta_{AB} + \zeta~\sum_{i=1}^N z_i~\Omega_{iA}~\Omega_{iB}
\end{eqnarray}
In the $\zeta\rightarrow\infty$ limit, which (with some care) can be thought of as a $1/z_i\rightarrow 0$ limit, we have
\begin{eqnarray}
 &&\Theta^{-1} = Z - Z~\Omega~Q^{-1}~\Omega^T~Z\\
 &&Q \equiv \Omega^T~Z~\Omega\\
 &&{\widetilde R} = \Theta^{-1}~R
\end{eqnarray}
where ${\widetilde R}$ is the vector of regressed returns in (\ref{w.reg.4}). So, regression is indeed a limit of optimization where the covariance matrix is given by $\Theta$. This is the factor model form -- with a subtlety, that is (see below).

\subsection{Factor Models}

{}In a multi-factor risk model, instead of $N$ stock returns $R_i$, one deals with $K\ll N$ risk factors and the covariance matrix $C_{ij}$ is replaced by $\Theta_{ij}$ given by
\begin{eqnarray}\label{Gamma}
 &&\Theta \equiv \Xi + {\widetilde \Omega}~\Phi~{\widetilde \Omega}^T\\
 && \Xi_{ij} \equiv \xi_i^2 ~\delta_{ij}
\end{eqnarray}
where $\xi_i$ is the specific (a.k.a. idiosyncratic) risk for each stock; ${\widetilde \Omega}_{iA}$ is an $N\times K$ factor loadings matrix; and $\Phi_{AB}$ is the factor covariance matrix, $A,B=1,\dots,K$. {\em I.e.}, the random processes $\Upsilon_i$ corresponding to $N$ stocks are modeled via $N$ random processes $\chi_i$ (corresponding to specific risk) together with $K$ random processes $f_A$ (corresponding to factor risk):
\begin{eqnarray}
 &&\Upsilon_i = \chi_i + \sum_{A=1}^F {\widetilde \Omega}_{iA}~f_A\\
 &&\left<\chi_i, \chi_j\right> = \Xi_{ij}\\
 &&\left<\chi_i, f_A\right> = 0\\
 &&\left<f_A, f_B\right> = \Phi_{AB}\\
 &&\left<\Upsilon_i, \Upsilon_j\right> = \Theta_{ij}
\end{eqnarray}
Instead of an $N \times N$ covariance matrix $C_{ij}$ we now have a $K \times K$ factor covariance matrix $\Phi_{AB}$. We have
\begin{eqnarray}
 &&\Theta = \Xi + \Omega~\Omega^T \\
 &&\Omega \equiv {\widetilde \Omega}~{\widetilde \Phi}\\
 &&{\widetilde \Phi}~{\widetilde \Phi}^T = \Phi
\end{eqnarray}
where ${\widetilde \Phi}_{AB}$ is the Cholesky decomposition of $\Phi_{AB}$, which is assumed to be positive-definite. Note that, in the notations of the previous subsection, we have chosen the normalization such that $\zeta = 1$.

{}In the factor model approach, one replaces the sample covariance matrix $C_{ij}$ (which is computed based on the time series of the returns $R_i$) by $\Theta_{ij}$. The main reason for doing so is that the off-diagonal elements of $C_{ij}$ typically are not expected to be too stable out-of-sample. In this regard, a constructed factor model covariance matrix $\Theta_{ij}$ is expected to be much more stable. This is because the number of factors, for which the factor covariance matrix $\Phi_{AB}$ needs to be computed, is $K\ll N$. Furthermore, if $M<N$ (recall that $M+1$ is the number of observations in each time series), then $C_{ij}$ is singular -- it has only $M<N$ nonzero eigenvalues in this case. Note that, assuming all specific risks $\xi_i>0$ and the factor covariance matrix $\Phi_{AB}$ is positive-definite, then $\Theta_{ij}$ is automatically positive-definite (and invertible).

\subsection{Optimization with Factor Model}

{}So, suppose we have a factor model covariance matrix $\Theta_{ij}$. If we maximize the Sharpe ratio using this factor model covariance matrix, the resulting holding weights are given by ($\lambda$ is fixed via (\ref{w.norm}))
\begin{eqnarray}
 &&w_i = {1\over \lambda~\xi^2_i}~\left(R_i - \sum_{j = 1}^N {R_j \over \xi^2_j}~\sum_{A,B = 1}^K \Omega_{iA}~\Omega_{jB}~{\widetilde Q}^{-1}_{AB} \right)\\
 &&{\widetilde Q}_{AB}\equiv \delta_{AB} + \sum_{i = 1}^N {1\over \xi^2_i}~\Omega_{i A}~\Omega_{i B}
\end{eqnarray}
where ${\widetilde Q}^{-1}_{AB}$ is the inverse of ${\widetilde Q}_{AB}$. As in the general case, these holding weights are not dollar neutral.

\subsubsection{Linear Constraints}\label{sub2.7.2}

{}As in the general case, in the factor model context too we can incorporate multiple (homogeneous) linear constrains (\ref{opt.mu.7}). Let ${\widehat \Omega}_{i\alpha}$, $\alpha\in H\equiv\{a\}\cup\{A\}$ ({\em i.e.}, the index $\alpha$ has $m$ values corresponding to the index $a$ and $K$ values corresponding to the index $A$) be the following $N\times (K+m)$ matrix:
\begin{eqnarray}
 &&{\widehat \Omega}_{ia} \equiv Y_{ia}\\
 &&{\widehat \Omega}_{iA} \equiv \Omega_{iA}
\end{eqnarray}
The corresponding holding weights then are given by:
\begin{eqnarray}
 &&w_i = {1\over \lambda~\xi^2_i}~\left(R_i - \sum_{j = 1}^N {R_j \over \xi^2_j}~\sum_{\alpha,\beta\in H} {\widehat \Omega}_{i\alpha}~{\widehat \Omega}_{j\beta}~{\widehat Q}^{-1}_{\alpha\beta} \right)\\
 &&{\widehat Q}_{\alpha\beta} \equiv \varphi_{\alpha\beta} + \sum_{i=1}^N{1\over\xi_i^2}~{\widehat \Omega}_{i\alpha}~{\widehat \Omega}_{i\beta}
\end{eqnarray}
where ${\widehat Q}^{-1}_{\alpha\beta}$ is the inverse of ${\widehat Q}_{\alpha\beta}$, and $\varphi_{AB} = \delta_{AB}$, $\varphi_{ab} = \varphi_{aA} = \varphi_{Aa} = 0$. We have
\begin{equation}
 \sum_{i=1}^N w_i~Y_{ia} = \sum_{i=1}^N w_i~{\widehat \Omega}_{ia}= \sum_{\alpha,\beta\in H}
 \sum_{j=1}^N{R_j\over\xi_j^2}~{\widehat\Omega}_{j\beta}~\varphi_{a\alpha}~{\widehat Q}^{-1}_{\alpha\beta} = 0
\end{equation}
So, the holding weights $w_i$ satisfy the constraints (\ref{opt.mu.7}).

\subsubsection{Optimization with Constraints}\label{sub2.9}

{}The constraints (\ref{opt.mu.7}) typically are related to risk management. Apart from dollar neutrality ({\em i.e.}, roughly, the market neutrality constraint), other constraints typically are the requirements of neutrality w.r.t. other risk factors, {\em e.g.}, industry neutrality, neutrality w.r.t. style risk factors ({\em e.g.}, size, liquidity, volatility, momentum, {\em etc.}) or other non-industry risk factors ({\em e.g.}, principal component based risk factors or betas). In practice, one often uses the same risk factors in $Y_{ia}$ as those in the factor loadings matrix $\Omega_{iA}$ in the factor model.\footnote{\, More precisely, usually one would use the unrotated factor loadings ${\widetilde \Omega}_{iA}$ -- recall that $\Omega = {\widetilde \Omega}~{\widetilde\Phi}$, where ${\widetilde \Phi}$ is the Cholesky decomposition of the factor covariance matrix $\Phi$. However, a rotation $Y \rightarrow Y U$ by an arbitrary nonsingular $m\times m$ matrix $U_{ab}$ does not change the constraints (\ref{opt.mu.7}).\label{foot.rot}} If that is the case, then there is certain redundancy in the matrix ${\widehat \Omega}_{i\alpha}$, which we turn to next.

{}Since we can always rotate the constraints (\ref{opt.mu.7}) by an arbitrary non-singular $m\times m$ matrix, we can separate these constraints into two sets, $\{a\} = \{a^\prime\}\cup\{a^{\prime\prime}\} \equiv J^\prime\cup J^{\prime\prime}$, such that $Y_{ia^{\prime\prime}}$ are ``orthogonal" to $\Omega_{iA}$ and no further rotation can make $Y_{ia^{\prime}}$ ``orthogonal" to $\Omega_{iA}$:
\begin{equation}
 \sum_{i=1}^N{1\over\xi_i^2}~\Omega_{iA}~Y_{ia^{\prime\prime}} = 0,~~~A=1,\dots,K,~~~a^{\prime\prime}\in J^{\prime\prime}
\end{equation}
Let us assume $J^{\prime\prime}$ is not empty -- if it is empty, we can still proceed as below, except that $\epsilon_i^{\prime\prime} = R_i$ in this case (see below).

{}Let $H^\prime \equiv\{A\}\cup J^\prime = H \setminus J^{\prime\prime}$ (recall that $H = \{A\} \cup \{a\}$). Then we have
\begin{eqnarray}
 w_i &=& {1\over \lambda~\xi^2_i}~\left(R_i - \sum_{j = 1}^N {R_j \over \xi^2_j}~\sum_{\alpha,\beta\in H} {\widehat \Omega}_{i\alpha}~{\widehat \Omega}_{j\beta}~{\widehat Q}^{-1}_{\alpha\beta} \right) =\nonumber\\
 &=& {1\over \lambda~\xi^2_i}~\left(\varepsilon^{\prime\prime}_i - \sum_{j = 1}^N {\varepsilon^{\prime\prime}_j \over \xi^2_j}~\sum_{\alpha^\prime,\beta^\prime\in H^\prime} {\widehat \Omega}_{i\alpha^\prime}~{\widehat \Omega}_{j\beta^\prime}~{\widehat Q}^{-1}_{\alpha^\prime\beta\prime} \right)
\end{eqnarray}
where
\begin{eqnarray}
 &&\varepsilon^{\prime\prime}_i \equiv R_i - \sum_{j = 1}^N {R_j \over \xi^2_j}~\sum_{a^{\prime\prime},b^{\prime\prime}\in J^{\prime\prime}} Y_{ia^{\prime\prime}}~Y_{jb^{\prime\prime}}~Q^{-1}_{a^{\prime\prime}b^{\prime\prime}}\\
 &&Q_{a^{\prime\prime}b^{\prime\prime}}\equiv\sum_{i=1}^N{1\over\xi_i^2}~Y_{ia^{\prime\prime}}~Y_{ib^{\prime\prime}}
\end{eqnarray}
and $Q^{-1}_{a^{\prime\prime}b^{\prime\prime}}$ is the inverse of the $\left|J^{\prime\prime}\right|\times \left|J^{\prime\prime}\right|$ matrix $Q_{a^{\prime\prime}b^{\prime\prime}}$, $a^{\prime\prime},b^{\prime\prime}\in J^{\prime\prime}$.

{}So, $\varepsilon^{\prime\prime}_i$ are nothing but the regression residuals of $R_i$ regressed over $Y_{ia^{\prime\prime}}$ with regression weights $z_i^{\prime\prime}\equiv 1/\xi_i^2$. Put differently, our original constrained optimization has reduced to constrained optimization with a subset of the original constraints
\begin{equation}
 \sum_{i=1}^N w_i~Y_{ia^\prime} = 0,~~~a^\prime\in J^\prime
\end{equation}
but instead of optimizing the returns $R_i$, we are now optimizing the regression residuals $\varepsilon^{\prime\prime}_i$. This is because the original matrix ${\widehat Q}_{\alpha\beta}$ is block-diagonal:
\begin{eqnarray}
 &&{\widehat Q}_{\alpha^\prime a^{\prime\prime}} = 0,~~~\alpha^\prime\in H^\prime,~a^{\prime\prime}\in J^{\prime\prime}\\
 &&{\widehat Q}_{a^{\prime\prime}b^{\prime\prime}} = Q_{a^{\prime\prime}b^{\prime\prime}},~~~a^{\prime\prime},b^{\prime\prime}\in J^{\prime\prime}
\end{eqnarray}
In fact, we can break this down further.

{}Let us assume that the $\left|J^\prime\right|$ columns in the remaining loadings $Y_{ia^\prime}$, $a^\prime\in J^\prime$ are a subset of the columns in the factor loadings $\Omega_{iA}$. So we have $\{A\} = \{A^\prime\} \cup J^\prime \equiv F^\prime \cup J^\prime$, and the $K^\prime \equiv \left|F^\prime\right| = K - \left|J^\prime\right|$ values of the index $A^\prime$ run over the columns in $\Omega_{iA}$ which differ from those in $Y_{ia^\prime}$. Further, to avoid notational confusion, we will denote $\left.\Omega_{iA}\right|_{A=A^\prime} \equiv \Omega^\prime_{iA^\prime}$, $A^\prime\in F^\prime$. It is then not difficult to show that
\begin{equation}
 {\widehat Q}^{-1}_{\alpha^\prime \beta^\prime}=\left( \begin{array}{lll}
               D^{-1}~~~ & \mathbb{O}~~~ & -D^{-1}E\Delta^{-1} \\
               \mathbb{O}~~~ & \mathbb{I}~~~ & -\mathbb{I}\\
               -\Delta^{-1}E^TD^{-1}~~~ & -\mathbb{I}~~~ & \mathbb{I} + \Delta^{-1} + \Delta^{-1}E^T D^{-1} E\Delta^{-1}\\
               \end{array}
        \right)
\end{equation}
Here $\mathbb{I}$ is the $\left|J^\prime\right|\times \left|J^\prime\right|$ identity matrix, while $\mathbb{O}$ is the $\left|F^\prime\right|\times \left|J^\prime\right|$ nil matrix, $D$ is an $\left|F^\prime\right|\times\left|F^\prime\right|$ matrix, $E$ is an $\left|F^\prime\right|\times \left|J^\prime\right|$ matrix and $\Delta$ is a $\left|J^\prime\right|\times \left|J^\prime\right|$ matrix:
\begin{eqnarray}
 && D \equiv {\widetilde Q}^\prime - E\Delta^{-1}E^T\\
 && {\widetilde Q}^\prime_{A^\prime B^\prime} \equiv \delta_{A^\prime B^\prime} + \sum_{i=1}^N{1\over\xi_i^2}~\Omega^\prime_{iA^\prime}~\Omega^\prime_{iB^\prime},~~~
 A^\prime,B^\prime \in F^\prime\\
 &&E_{A^\prime b^\prime} \equiv\sum_{i=1}^N {1\over\xi_i^2}~\Omega^\prime_{iA^\prime}~Y_{ib^\prime},~~~
 A^\prime \in F^\prime,~~~b^\prime\in J^\prime\\
 &&\Delta_{a^\prime b^\prime}\equiv \sum_{i=1}^N {1\over\xi_i^2}~Y_{ia^\prime}~Y_{ib^\prime},~~~
 a^\prime,b^\prime\in J^\prime
\end{eqnarray}
We therefore have (in matrix notation -- here $Y$ refers to the $Y_{ia^\prime}$ matrix)
\begin{eqnarray}
 w &=& {1\over\lambda}\left\{\Xi^{-1} - \Xi^{-1}\left[\Omega^\prime D^{-1}\Omega^\prime - \Omega^\prime D^{-1}E\Delta^{-1}Y^T - Y\Delta^{-1}E^TD^{-1}\Omega^\prime\right.\right. + \nonumber\\
 &+&\left.\left.Y\left(\Delta^{-1} + \Delta^{-1}E^TD^{-1}E\Delta^{-1}\right)Y^T
 \right]\Xi^{-1}\right\}\varepsilon^{\prime\prime}\label{w.opt.gen}
\end{eqnarray}
Furthermore
\begin{equation}
 Y^T~w = 0
\end{equation}
In fact, $w_i$ given by (\ref{w.opt.gen}) correspond to optimizing the residuals $\varepsilon^{\prime\prime}_i$ using a reduced factor model with the same specific risk but the factor loadings given by $\Omega^\prime_{iA^\prime}$
\begin{equation}
 \Theta^\prime_{ij} \equiv \xi_i^2~\delta_{ij} + \sum_{A^\prime=1}^{K^\prime} \Omega^\prime_{iA^\prime}~\Omega^\prime_{jA^\prime}
\end{equation}
subject to the constraints
\begin{equation}
 \sum_{i=1}^N w_i~Y_{ia^\prime} = 0,~~~a^\prime\in J^\prime
\end{equation}
The solution to this optimization problem is given by
\begin{equation}\label{w.opt.gen.1}
 w_i = {1\over \lambda~\xi^2_i}~\left(\varepsilon^{\prime\prime}_i - \sum_{j = 1}^N {\varepsilon^{\prime\prime}_j \over \xi^2_j}~\sum_{\alpha^*,\beta^* \in F^*} {\widehat \Omega}_{i\alpha^*}~{\widehat \Omega}_{j\beta^*}~{\widehat Q}^{-1}_{\alpha^*\beta^*} \right)
\end{equation}
where $F^*\equiv F^\prime\cup J^\prime$ (so $F^*$ as a set is the same as $\{A\}$, but we use a different notation for it to avoid confusion), and we have $\left.{\widehat\Omega}_{i\alpha^*}\right|_{\alpha^* = A^\prime} \equiv\Omega^\prime_{iA^\prime}$, $\left.{\widehat\Omega}_{i\alpha^*}\right|_{\alpha^* = a^\prime} \equiv Y_{ia^\prime}$, and
\begin{equation}
 {\widehat Q}^{-1}_{\alpha^* \beta^*}=\left( \begin{array}{ll}
               D^{-1}~~~ & -D^{-1}E\Delta^{-1} \\
               -\Delta^{-1}E^TD^{-1}~~~ & \Delta^{-1} + \Delta^{-1}E^T D^{-1} E\Delta^{-1}\\
               \end{array}
        \right)
\end{equation}
It then follows that $w_i$ in (\ref{w.opt.gen}) and (\ref{w.opt.gen.1}) are identical.

{}To summarize, optimization is done with the columns in the factor loadings matrix $\Omega_{iA}$ corresponding to the columns in $Y_{ia}$ omitted.

\subsection{Pitfalls}

{}So, what happens if we run constrained optimization with the factor loadings matrix as $Y_{ia}$? {\em I.e.}, $m=K$, the index $a$ takes the same values as the index $A$, and $\left.Y_{ia}\right|_{a = A} = \Omega_{iA}$. (If we wish to have dollar neutrality, we simply assume that $\Omega_{iA}$ contains the intercept.) In this case, using the results of the previous subsection:
\begin{eqnarray}
 &&w_i = {1\over \lambda~\xi^2_i}~\left(R_i - \sum_{j = 1}^N {R_j \over \xi^2_j}~\sum_{A,B=1}^K \Omega_{iA}~\Omega_{iB}~Q^{-1}_{AB} \right)\label{w.reg.spec.risk}\\
 &&Q_{AB}\equiv\sum_{i=1}^N{1\over\xi_i^2}~\Omega_{iA}~\Omega_{iB}
\end{eqnarray}
so $w_i$ are the same as in the weighted regression with regression weights $z_i = 1/\xi_i^2$.

\subsubsection{Specific Risk or Total Risk?}

{}In the optimization context, the regression weights in (\ref{w.reg.spec.risk}) naturally come out to be $z_i = 1/\xi_i^2$, the inverse of specific volatility squared, not total volatility ({\em i.e.}, $z_i\not=1/\sigma_i^2$). This addresses the subtlety mentioned at the end of Subsection \ref{sub2.5}. However, {\em a priori} there is nothing wrong with using $z_i = 1/\sigma_i^2$ in the weighted regression outside of the factor model context. Specific risk is not known unless a factor model is available or carefully constructed. In that case, total risk is what is available for using as the regression weights, and typically can be so used.

\subsubsection{Optimization of Regression Residuals}

{}Instead of imposing constraints in optimization, one may be tempted to first regress returns $R_i$ over (some) factor loadings $\Omega_{iA}$ to obtain regression residuals $\varepsilon_i$, and do the optimization based on these residuals (as opposed to the returns $R_i$). The rationale here is that the regressed returns ${\widetilde R}_i = z_i~\varepsilon_i$ (where $z_i$ are the regression weights) are neutral w.r.t. the loadings used in the regression.
However, unless this is done correctly, the resulting holding weights will not be neutral w.r.t. $\Omega_{iA}$ as the optimization in general undoes any such neutrality.

{}Thus, consider the following strategy:
\begin{eqnarray}\label{reg.res.opt}
 &&\varepsilon_i \equiv R - Y~(Y^T~Z~Y)^{-1}~Y^T~Z~R\\
 &&Z \equiv \mbox{diag}(z_i)\\
 &&w_i \equiv {1\over \lambda~\xi^2_i}~\left(\varepsilon_i - \sum_{j = 1}^N {\varepsilon_j \over \xi^2_j}~\sum_{A,B=1}^K \Omega_{iA}~\Omega_{jB}~{\widetilde Q}^{-1}_{AB} \right)\\
 &&{\widetilde Q}_{AB} \equiv \delta_{AB} + \sum_{i=1}^N {1\over\xi_i^2}~\Omega_{iA}~\Omega_{iB}
\end{eqnarray}
Here we have purposefully kept the loadings $Y_{ia}$ in the weighted regression (first equation above) distinct from the factor loadings $\Omega_{iA}$ in the optimization (third equation above). Note that the optimization is done on the regression residuals $\varepsilon_i$, not on the returns $R_i$.

{}If the purpose of the regression is that the holding weights be neutral w.r.t. $Y_{ia}$, then to ensure this, the matrix $Y_{ia}$ must be the same as $\Omega_{iA}$ (modulo immaterial rotations -- see footnote \ref{foot.rot}), {\em i.e.}, $m = K$, and the regression weights $z_i$ cannot be arbitrary but must be taken as inverse specific variances: $z_i = 1/\xi_i^2$. Indeed, from (\ref{reg.res.opt}) we then have
\begin{eqnarray}
 &&\sum_{i=1}^N {\widetilde R}_i~\Omega_{iA} = 0\\
 &&{\widetilde R}_i \equiv z_i~\varepsilon_i = {1\over\xi_i^2}~\varepsilon_i
\end{eqnarray}
so that
\begin{eqnarray}
 &&w_i = {1\over\lambda}~{\widetilde R}_i\\
 &&\sum_{i=1}^N w_i~\Omega_{iA} = 0
\end{eqnarray}
{\em I.e.}, optimization of regression residuals (up to an overall proportionality constant $1/\lambda$) simply reduces to the regressed returns ${\widetilde R}_i$, which are neutral w.r.t. $\Omega_{iA}$.

\section{``Intermezzo"}

{}In Section \ref{sec.mean.rev} we started with simple pair trading and by the end of the last section it got substantially more involved. This trend is going to continue in the following sections, so this is a good place for an ``intermezzo". We will try to keep it light.

{}So, consider two stocks, A and B. Let their (sample) covariance matrix be
\begin{equation}\label{CAB}
 C = \left( \begin{array}{ll}
               \sigma_A^2~~~ & \rho~\sigma_A~\sigma_B\\
               \rho~\sigma_A~\sigma_B~~~ & \sigma_B^2 \\
               \end{array}\right)
\end{equation}
Here $\sigma_A$ and $\sigma_B$ are the volatilities, and $\rho$ is the correlation. Let our portfolio have $D_A$ and $D_B$ dollar holdings in A and B. Let $R_A$ and $R_B$ be the expected returns for A and B. Then the expected Sharpe ratio of this portfolio is
\begin{equation}
 S = {{D_A~R_A + D_B~R_B}\over\sqrt{(\sigma_A~D_A)^2 + (\sigma_B~D_B)^2 +2~\rho~(\sigma_A~D_A)(\sigma_B~D_B)}}
\end{equation}
It is maximized by
\begin{eqnarray}
 &&D_A = \gamma\left({R_A\over\sigma^2_A} - {\rho~R_B\over\sigma_A~\sigma_B}\right)\\
 &&D_B = \gamma\left({R_B\over\sigma^2_B} - {\rho~R_A\over\sigma_A~\sigma_B}\right)
\end{eqnarray}
Here $\gamma>0$ is an arbitrary constant, which is a consequence of the invariance of $S$ under simultaneous rescalings $D_A\rightarrow\zeta D_A$, $D_B\rightarrow\zeta D_B$ ($\zeta > 0$), and it is fixed via the requirement that $|D_A|+|D_B| = I$, where $I$ is the investment level.

{}Now let us assume that the volatilities are the same: $\sigma_A = \sigma_B\equiv\sigma$. We have
\begin{eqnarray}
 &&D_A = {\gamma\over\sigma^2} \left(R_A - \rho~R_B\right)\\
 &&D_B = {\gamma\over\sigma^2} \left(R_B - \rho~R_A\right)
\end{eqnarray}
As $\rho\rightarrow 1$, we have $D_A + D_B \rightarrow 0$, which is the dollar neutrality condition. So, the $S\rightarrow \mbox{max}$ optimization, in the limit where volatilities are identical and the correlation goes to 1, produces a dollar neutral portfolio. How come? Two answers.

{}First, when $\sigma_A=\sigma_B$ and $\rho=1$, the covariance matrix $C$ is singular. The eigenvector $V_i$ corresponding to the null eigenvalue is $V^T = (1,~-1)$, and in this direction the portfolio volatility vanishes and the Sharpe ratio goes to infinity (see below). This is why $D_B = -D_A$ maximizes the Sharpe ratio.\footnote{\,For $\rho=\pm 1$ the Sharpe ratio goes to infinity if $R_A\neq \pm R_B$: the volatility vanishes for $D_B=\mp D_A$ (recall that $\sigma_A = \sigma_B$). If $R_A=\pm R_B$, then the two instruments $A$ and -- long for plus sign and short for minus sign -- $B$ are indistinguishable for optimization purposes.}

{}Second, we can tie this to Subsection \ref{sub2.5}. Consider a {\em one}-factor model for two stocks $A$ and $B$: $\Theta = \Xi + \zeta~\Omega~\Omega^T$, where   $\Xi = \mbox{diag}(\xi_A^2,~\xi_B^2)$, and $\Omega^T = (1,~1)$. Then $\Theta = C$ in (\ref{CAB}) with $\sigma_{A,B}^2 = \xi_{A,B}^2 + \zeta$, and $\rho = \zeta / (\sigma_A~\sigma_B)$. In the $\zeta\rightarrow\infty$ limit (with $\xi_A$ and $\xi_B$ fixed) we have $\sigma_{A,B}^2\rightarrow \zeta\equiv \sigma^2$, and $\rho\rightarrow 1$, exactly as above. On the other hand, as we saw in Subsection \ref{sub2.5}, in this limit optimization reduces to a regression over $\Omega$, which is nothing but the intercept, hence dollar neutrality.

\section{Optimization with Costs}\label{sec.opt.cost}

\subsection{Linear Costs}

{}Above we ignored trading costs. Let us start by adding linear costs:\footnote{\, For the sake of simplicity, the transaction costs for buys and sells are assumed to be the same.}
\begin{equation}
 P = \sum_{i=1}^N R_i~D_i - \sum_{i=1}^N {\widetilde L}_i~\left|D_i - D_i^*\right|
\end{equation}
where ${\widetilde L}_i$ for each stock includes, per each {\em dollar} traded, all fixed trading costs (SEC fees, exchange fees, broker-dealer fees, {\em etc.}) and linear slippage. The linear cost assumes no impact, {\em i.e.}, trading does not affect the stock prices. Also, $D_i$ are the desired dollar holdings, and $D_i^*$ are the {\em current} dollar holdings. For the purposes of optimization, as above, it is more convenient to deal with the holding weights $w_i$ instead of the dollar holdings $D_i$. Let ${\widetilde L}_i \equiv I~L_i$, $D_i\equiv I~w_i$ and $D_i^*\equiv I~w_i^*$. Then
\begin{equation}
 {\widetilde P} = {P \over I} = \sum_{i=1}^N R_i~w_i - \sum_{i=1}^N L_i~\left|w_i - w_i^*\right|
\end{equation}
As above, we can have constraints
\begin{equation}\label{const.w.s}
 \sum_{i=1}^N w_i~Y_{ia} = 0,~~~a=1,\dots,m
\end{equation}
We will assume that the current holdings satisfy the same constraints:
\begin{equation}\label{const.w.star}
 \sum_{i=1}^N w_i^*~Y_{ia} = 0,~~~a=1,\dots,m
\end{equation}
This includes establishing trades ($w_i^*\equiv 0$). We have the normalization condition (\ref{w.norm}) for $w_i$, but not necessarily for $w_i^*$ ({\em e.g.}, if the position is being established).

\subsection{Optimization with Costs and Homogeneous Constraints}

{}More generally, costs can be modeled by some function $f(w)$ of $w_i$, which also depends on the current holding weights $w_i^*$, but the precise form of this dependence is not going to be important here. We have
\begin{equation}
 {\widetilde P} = \sum_{i=1}^N R_i~w_i - f
\end{equation}
Generally, the costs spoil the invariance of the Sharpe ratio
\begin{equation}
 S = {{\widetilde P}\over {\widetilde V}} = {{\sum_{i=1}^N R_i~w_i - f}\over\sqrt{{\sum_{i=1}^N C_{ij}~w_i~w_j}}}
\end{equation}
under the rescaling $w_i \rightarrow \zeta w_i$ ($\zeta > 0$) with a single exception of $f(w) = f_{\rm{\scriptstyle{special}}}(w)$:
\begin{equation}\label{special.app}
 f_{\rm{\scriptstyle{special}}}(w) \equiv \sum_{i=1}^N L_i~\left|w_i\right|
\end{equation}
where $L_i$ are positive constants. The costs are of this form when we have only linear costs and the current holdings are zero, {\em i.e.}, this is an establishing trade. Below we will assume that $f(w)$ is {\em not} of this form.

{}In the absence of the rescaling invariance, care is needed when rewriting the Sharpe ratio maximization problem in terms of minimizing an objective function. The Sharpe ration maximization problem reads:
\begin{equation}\label{S.L.app}
 {\widetilde S} \equiv S + \sum_{a=1}^m\sum_{i=1}^N w_i~Y_{ia}~\mu_a + {\widetilde \mu}\left(\sum_{i=1}^N\left|w_i\right| - 1\right)
\end{equation}
We need to maximize ${\widetilde S}$ w.r.t. $w_i$ and Lagrange multipliers $\mu_a$ and ${\widetilde \mu}$, which gives:\footnote{\, Actually, $w_i$ derivatives are defined only for $w_i \neq 0$ and, {\em e.g.}, in the case of linear costs for $w_i \neq w_i^*$ -- see Subsection \ref{sub4.1} for details.}
\begin{eqnarray}\label{w.eq.s.app}
 && {1\over {\widetilde V}}\left[R_i - f_i - \lambda\sum_{j=1}^N C_{ij}~w_j\right] +\sum_{a=1}^m Y_{ia}~\mu_a +
 {\widetilde \mu}~\mbox{sign}\left(w_i\right) = 0\\
 &&\sum_{i=1}^N w_i~Y_{ia} = 0\label{const.5}\\
 &&\sum_{i=1}^N\left|w_i\right| = 1\label{norm.5}\\
 &&\lambda\equiv {{\sum_{i=1}^N R_i~w_i - f}\over{\sum_{i=1}^N C_{ij}~w_i~w_j}}\label{lambda.s.app}\\
 &&f_i \equiv {\partial f\over\partial w_i}
\end{eqnarray}
If we multiply the first equation by $w_i$ and sum over $i$, we get
\begin{equation}
 {\widetilde\mu} = {1\over {\widetilde V}}\left[\sum_{i=1}^Nf_i~w_i-f\right]
\end{equation}
Unless $f(w)$ has the special form (\ref{special.app}), which we assume not to be the case, then generally ${\widetilde\mu}\neq 0$.

{}We can still {\em formally} recast the Sharpe ratio maximization in terms of minimizing the following objective function (w.r.t. $w_i$ and Lagrange multipliers $\mu^\prime_a$ and ${\widetilde\mu}^\prime$):
\begin{eqnarray}\label{obj.right}
 && g(w,\mu^\prime,{\widetilde\mu}^\prime, \lambda^\prime) \equiv {\lambda^\prime\over 2}\sum_{i,j=1}^N C_{ij}~w_i~ w_j - \sum_{i=1}^N R_i~w_i + f -\nonumber\\
 &&\,\,\,\,\,\,\,
 -\sum_{a=1}^m\sum_{i=1}^N w_i~Y_{ia}~\mu^\prime_a - {\widetilde \mu}^\prime\left(\sum_{i=1}^N\left|w_i\right| - 1\right)\label{obj.right.app} \\
 && g(w,\mu^\prime, {\widetilde\mu}^\prime, \lambda^\prime)\rightarrow \mbox{min}
\end{eqnarray}
The minimization equations read:
\begin{eqnarray}\label{w.sol.app.prime}
 &&\lambda^\prime \sum_{j=1}^N C_{ij}~w_j - R_i + f_i - \sum_{a=1}^m Y_{ia}~\mu_a^\prime - {\widetilde \mu}^\prime~\mbox{sign}\left(w_i\right) = 0\\
 &&\sum_{i=1}^N w_i~Y_{ia} = 0\\
 &&\sum_{i=1}^N\left|w_i\right| = 1
\end{eqnarray}
Multiplying the first equation by $w_i$ and summing over $i$, we get
\begin{equation}
 {\widetilde\mu}^\prime = \lambda^\prime\sum_{i,j=1}^N C_{ij}~w_i~w_j - \sum_{i=1}^N \left[R_i - f_i\right] w_i
\end{equation}
The statement then is that there exists a value of $\lambda^\prime$ for which minimizing the objective function produces the same solution for $w_i$ as maximizing the Sharpe ratio. This value is given by $\lambda^\prime = \lambda$, where $\lambda$ is given by (\ref{lambda.s.app}) with $w_i$ corresponding to the {\em optimal solution}, {\em i.e.}, the maximal Sharpe ratio solution. We then have
\begin{eqnarray}
 &&\lambda^\prime = \lambda\\
 &&\mu^\prime_a = {\widetilde V}~\mu_a\\
 &&{\widetilde\mu}^\prime = {\widetilde V}~{\widetilde\mu} = \sum_{i=1}^N f_i~w_i - f \label{mu.prime}
\end{eqnarray}
However, the practical value of this statement is limited -- unless we solve the Sharpe ratio maximization problem, we do not know what $\lambda$ is. The Sharpe ratio maximization problem is highly nonlinear and prone to usual nonlinear instabilities. On the other hand, in terms of minimizing the objective function, we can treat $\lambda^\prime$ as a parameter. Then the problem of maximizing the Sharpe ratio reduces to a one-dimensional problem of finding the value of $\lambda^\prime$ for which the Sharpe ratio is maximal.

\subsubsection{Pitfalls}\label{appA2.1}

{}Because in the presence of costs the rescaling invariance is lost, the maximum Sharpe ratio solution is {\em not} given by the following minimization (w.r.t. $w_i$ and Lagrange multipliers $\mu_a^{\prime\prime}$):
\begin{eqnarray}\label{obj.wrong}
 && {\widetilde g}(w,\mu^{\prime\prime},\lambda^{\prime\prime}) \equiv {\lambda^{\prime\prime}\over 2}\sum_{i,j=1}^N C_{ij}~w_i~ w_j - \sum_{i=1}^N R_i~w_i + f - \nonumber\\
 &&\,\,\,\,\,\,\, -\sum_{a=1}^m\sum_{i=1}^N w_i~Y_{ia}~\mu^{\prime\prime}_a \\
 && {\widetilde g}(w,\mu^{\prime\prime}, \lambda^{\prime\prime})\rightarrow \mbox{min}
\end{eqnarray}
It is incorrect to assume -- and this appears to be a common misstep in practical applications -- that the maximum Sharpe ratio solution is given by the solution to this minimization condition for the value of $\lambda^{\prime\prime}$ such that (\ref{norm.5}) is satisfied.

{}To see this, for the sake of simplicity, let us assume that there are no linear constraints (\ref{const.5}). Then we have
\begin{equation}\label{w.eq.wrong}
 \lambda^{\prime\prime} \sum_{j=1}^N C_{ij}~w_j - R_i + f_i = 0
\end{equation}
Multiplying this equation by $w_i$ and summing over $i$, we get
\begin{equation}
 \lambda^{\prime\prime} = {{\sum_{i=1}^N \left[R_i - f_i\right] w_i}\over{\sum_{i,j=1}^N C_{ij}~w_i~w_j}}
\end{equation}
Then, for this solution to coincide with (\ref{w.sol.app.prime}) (without homogeneous constraints (\ref{const.5}), that is), we must have
\begin{equation}\label{w.bad.1}
 {1\over \sum_{i,j=1}^N C_{ij}~w_i~w_j}~\sum_{j=1}^N C_{ij}~w_j = \mbox{sign}\left(w_i\right)
\end{equation}
where we have taken into account that ${\widetilde \mu}^\prime \neq 0$ -- see (\ref{mu.prime}). Plugging (\ref{w.bad.1}) back into (\ref{w.eq.wrong}), we get
\begin{eqnarray}
 &&R_i - f_i = \gamma~\mbox{sign}\left(w_i\right)\\
 &&\gamma \equiv \sum_{i=1}^N \left[R_i - f_i\right] w_i
\end{eqnarray}
However, this is impossible to satisfy for a general form of $f(w)$. {\em E.g.}, in the case of linear costs
\begin{equation}\label{lin.cost.app}
 f(w) = f_{\rm{\scriptstyle{linear}}}(w) \equiv \sum_{i=1}^N L_i~\left|w_i - w_i^*\right|
\end{equation}
and we cannot have $R_i - L_i~\mbox{sign}\left(w_i - w_i^*\right) = \gamma~\mbox{sign}\left(w_i\right)$ for all $i=1,\dots,N$. Therefore, minimizing the objective function (\ref{obj.wrong}) does {\em not} produce the maximal Sharpe ratio solution. The correct objective function to minimize is (\ref{obj.right.app}), where $\lambda^\prime$ is treated as a parameter, whose value is fixed via a one-dimensional search algorithm such that the Sharpe ratio is maximized.

\subsubsection{Global {\em vs.} Local Optima}\label{appA2.2}

{}Assuming the cost function $f(w)$ is convex, the ``wrong" objective function (\ref{obj.wrong}) is convex w.r.t. $w_i$, so it has a unique local minimum. However, the correct objective function (\ref{obj.right}) is not necessarily convex. This is because the contribution due to the ${\widetilde\mu}^\prime$ term is convex if and only if ${\widetilde\mu}^\prime \leq 0$, which is not necessarily the case -- see (\ref{mu.prime}). If ${\widetilde\mu}^\prime > 0$, there can be multiple local minima further complicating the search for a global minimum. {\em E.g.}, in the case of linear costs (\ref{lin.cost.app}), we have ${\widetilde\mu}^\prime = \sum_{i=1}^N L_i~w_i^*~\mbox{sign}\left(w_i - w_i^*\right)$, which need not be negative.

\subsection{Maximizing Sharpe Ratio with Linear Costs}

{}As we saw in the previous subsection, in the presence of costs the Sharpe ratio maximization problem is highly nonlinear and may not even have a unique local minimum -- even for linear costs (\ref{lin.cost.app}), assuming some $w_i^*\neq 0$.

{}So, how is this optimization done in practice? Often it is done by simply taking the ``wrong" objective function (\ref{obj.wrong}) and iterating $\lambda^{\prime\prime}$ until (\ref{w.norm}) is satisfied.\footnote{\, For any $\lambda^{\prime\prime} > 0$, there is a unique optimum assuming $C_{ij}$ is positive-definite, and all $L_i \geq 0$.} As discussed above, this solution does not maximize the Sharpe ratio. In some cases, it could be a reasonable approximation though. Let us focus on linear costs. Let: i) $L_i$ be uniform, $L_i\equiv L$; ii) $w_i^*\rightarrow w_i$ be a rebalancing trade from a previously optimized solution with $\sum_{i=1}^N \left|w_i^*\right| = 1$; iii) our portfolio be dollar neutral, so that $\sum_{i=1}^N w_i^* = \sum_{i=1}^N w_i = 0$; iv) the number of stocks be large ($N\gsim 1000$); and v) there be diversification constraints in place, so $w_i$ are not larger than, say, low single digit percent.\footnote{\, We will discuss bounds below. Alternatively, one can squash returns to achieve the same.} If $\mbox{sign}(w_i^*)$ and $\mbox{sign}(w_i-w_i^*)$ ({\em i.e.}, current holding signs and desired trade signs) are not highly correlated, then $|{\widetilde\mu}^\prime| \ll L$ and its contribution to (\ref{w.sol.app.prime}) is small compared with the contribution due to the linear costs, so we can approximately ignore it. And neglecting the ${\widetilde \mu}^\prime$ contribution in (\ref{obj.right}) is the same as using (\ref{obj.wrong}).\footnote{\, This argument also goes through for partially establishing and liquidating trades with $\xi \lsim 1$, {\em i.e.}, $\xi\equiv \sum_{i=1}^N |w_i^*|$ need not be equal 1. Uniformity of $L_i$ can also be relaxed (with some care).}

{}With the above in mind, we can minimize the objective function (\ref{obj.wrong}) and fix $\lambda^{\prime\prime}$ via an iterative procedure until (\ref{w.norm}) is satisfied. This is what is done in most practical applications. This also avoids the issue of multiple local optima discussed in Subsection \ref{appA2.2} -- the objective function (\ref{obj.wrong}) is convex and has a unique local minimum. However, we emphasize: this is only an {\em approximation} to Sharpe $\rightarrow$ max.

\section{Optimization: Costs, Constraints \& Bounds}\label{sec.opt}

{}So, let us consider the following optimization problem:
%\footnote{\, We are simplifying the notation of the previous section to get rid of extra primes, {\em etc.}}
\begin{eqnarray}
 && g(w,\mu, \lambda) \equiv {\lambda\over 2}\sum_{i,j=1}^N C_{ij} ~w_i ~w_j - \sum_{i=1}^N \left(R_i ~w_i - L_i\left|w_i-w_i^*\right|\right) -
 \nonumber\\
 &&\,\,\,\,\,\,\, -\sum_{a=1}^m\sum_{i=1}^N w_i~Y_{ia}~\mu_a \\
 && g(w,\mu, \lambda)\rightarrow \mbox{min}\\
 &&w^-_i \leq w_i \leq w^+_i\label{bounds}
\end{eqnarray}
where: the minimization is w.r.t. $w_i$ and Lagrange multipliers $\mu_a$; $\lambda$ is treated as a parameter to be fixed iteratively so that the normalization condition
\begin{equation}
 \sum_{i=1}^N \left|w_i\right| = 1
\end{equation}
is satisfied; and we have included lower $w^-_i$ and upper $w^+_i$ bounds (\ref{bounds}) on the holding weights. If there are no bounds, we can simply take $w^-_i$ and $w^+_i$ to be large negative and large positive numbers, respectively.

{}In the following it will be more convenient to use
\begin{equation}
 x_i \equiv w_i - w_i^*
\end{equation}
We then have
\begin{eqnarray}\label{obj.x}
 && {\widetilde g}(x,\mu, \lambda) \equiv {\lambda\over 2}\sum_{i,j=1}^N C_{ij} x_i x_j - \sum_{i=1}^N \left(\rho_i x_i - L_i\left|x_i\right|\right) -
 \sum_{a=1}^m\sum_{i=1}^N x_i Y_{ia} \mu_a \\
 && {\widetilde g}(x,\mu, \lambda)\rightarrow \mbox{min}\\
 &&x^-_i \leq x_i \leq x^+_i\label{bounds.x}\\
 &&\rho_i \equiv R_i - \lambda~\sum_{j=1}^N C_{ij}~w_j^*\\
 &&x^\pm_i \equiv w^\pm_i - w^*_i
\end{eqnarray}
Furthermore, we are assuming that the current holdings satisfy the linear constraints
\begin{equation}
 \sum_{i=1}^N w^*_i~Y_{ia} = 0,~~~a=1,\dots,m
\end{equation}
and we have dropped an immaterial constant term from the objective function (\ref{obj.x}).

\subsection{Bounds}\label{sub4.3}

{}Typically, in practical applications, bounds are used to cap i) the positions of individual stocks in a portfolio, and ii) the amount of trading in each stock. {\em E.g.}, let us assume that we impose the following constraints:
\begin{eqnarray}
 &&\left|w_i\right| \leq \xi~{\widetilde v}_i\\
 &&\left|w_i-w_i^*\right| \leq {\widetilde \xi}~{\widetilde v}_i\\
 &&{\widetilde v}_i \equiv {v_i\over I}
\end{eqnarray}
where $v_i$ is, say, a 20-day average daily dollar volume for the stock labeled by $i$, and $\xi$ and ${\widetilde \xi}$ are some positive percentages. Then we would have
\begin{eqnarray}
 &&x_i^+ = \mbox{min}\left({\widetilde \xi}~{\widetilde v}_i, ~\xi~{\widetilde v}_i - w_i^*\right) \geq 0\\
 &&x_i^- = \mbox{max}\left(-{\widetilde \xi}~{\widetilde v}_i, ~-\xi~{\widetilde v}_i - w_i^*\right)\leq 0
\end{eqnarray}
and we are assuming that $\left|w_i^*\right| \leq \xi~{\widetilde v}_i$. There is little to no value in trying to account for any isolated ``extraordinary" cases ({\em e.g.}, there is news for a given stock and it needs to be liquidated, which for $w_i^* > 0$ would mean that $x_i^+ < 0$, and for $w_i^* < 0$ it would mean that $x_i^- > 0$) as they can be simply treated by setting the desired holdings for such few stocks and altogether excluding them from optimization of the remaining universe of stocks. We will therefore assume that $x_i^+\geq 0$ and $x_i^-\leq 0$.

{}We can avoid much notational headache if we further assume that $x^+_i > 0 $ and $x^-_i < 0$. There are cases where one may wish to set $x^+_i = 0$ or $x^-_i = 0$, {\em e.g.}, we cannot sell a stock due to a short-sale restriction (hard-to-borrow stock, {\em etc.}). However, instead of setting $x_i^+$ or $x_i^-$ strictly to zero, it is more practical to set it to a small positive or negative number instead ({\em e.g.}, within the desired precision tolerance). In the following we will assume that $x^+_i > 0 $ and $x^-_i < 0$.

\subsection{Optimization: General Case}\label{sub4.1}

{}Let us define the following subsets of the index $i=1,\dots,N$:
\begin{eqnarray}
 &&x_i\neq 0,~~~i\in J\\
 &&x_i = 0,~~~i\in J^\prime\label{x.prime}\\
 &&x_i = x_i^+ > 0,~~~i\in J^+\subset J\\
 &&x_i = x_i^- < 0,~~~i\in J^-\subset J\\
 &&{\overline J} \equiv J^+ \cup J^-\\
 &&{\widetilde J} \equiv J \setminus {\overline J}\\
 &&\eta_i \equiv \mbox{sign}\left(x_i\right),~~~i\in J
\end{eqnarray}
Note that, since the modulus has a discontinuous derivative, the minimization equations are not the same as setting first derivatives of ${\widetilde g}(x,\mu,\lambda)$ w.r.t. $x_i$ and $\mu_a$ to zero. More concretely, first derivatives w.r.t. $x_i$ are well-defined for $i\in J$, but not for $i\in J^\prime$, while the first derivatives w.r.t. $\mu_a$ are always well-defined. Furthermore, first derivatives w.r.t. $x_i$ for $i\in {\overline J}$ ({\em i.e.}, at the bounds) need not be zero. Let us therefore consider the global minimum condition:
\begin{equation}\label{JJ1}
 \left.{\widetilde g}(x^\prime, \mu^\prime, \lambda)\right|_{x^\prime_i = x_i + \epsilon_i,~\mu^\prime_a = \mu_a + \epsilon_a} \geq {\widetilde g}(x_i, \mu_a, \lambda)
\end{equation}
Here $\epsilon_i$ and $\epsilon_a$ {\em a priori} are arbitrary except that at the bounds we have
\begin{eqnarray}\label{e+}
 &&\epsilon_i \leq 0,~~~i\in J^+\\
 &&\epsilon_i \geq 0,~~~i\in J^-\label{e-}
\end{eqnarray}
{}From (\ref{JJ1}) we get
\begin{eqnarray}
 &&{\lambda\over 2}\sum_{i,j=1}^N C_{ij} \epsilon_i \epsilon_j + \sum_{i\in J}L_i\left(\left|x_i + \epsilon_i\right| - \left|x_i\right| - \eta_i \epsilon_i\right) - \sum_{a=1}^m\sum_{i=1}^N \epsilon_i Y_{ia} \epsilon_a - \nonumber\\
 &&\sum_{a=1}^m\sum_{i=1}^N x_i Y_{ia} \epsilon_a +\sum_{j=1}^N\left(\lambda \sum_{i\in J} C_{ij} x_i -\rho_j + L_j {\widetilde \eta}_j- \sum_{a=1}^m Y_{ja} \mu_a\right)\epsilon_j \geq 0\label{global.min}
\end{eqnarray}
where (the ambiguity in $\mbox{sign}(\epsilon_i)$ below is immaterial; we can set $\mbox{sign}(0)=0$)
\begin{eqnarray}
 &&{\widetilde\eta}_i \equiv \eta_i,~~~i\in J\\
 &&{\widetilde\eta}_i \equiv \mbox{sign}(\epsilon_i),~~~i\in J^\prime
\end{eqnarray}
The first line in (\ref{global.min}) is ${\cal O}(\epsilon^2)$. For infinitesimal $\epsilon_i$ the second line gives:
\begin{eqnarray}\label{J1.bounds}
 &&\lambda \sum_{j\in J} C_{ij}~x_j -\rho_i + L_i~\eta_i - \sum_{a=1}^m Y_{ia}~\mu_a = 0,~~~i\in {\widetilde J}\\
 &&\sum_{i=1}^N x_i~Y_{ia} = \sum_{i\in J}^N x_i~Y_{ia} = 0,~~~a=1,\dots,m\label{Y1.bounds}
\end{eqnarray}
which equations correspond to setting to zero first derivatives of ${\widetilde g}(x, \mu, \lambda)$ w.r.t. $x_i$, $i\in {\widetilde J}$ and $\mu_a$, and then we also have the following inequalities for $i\not\in {\widetilde J}$:
\begin{eqnarray}\label{globalmin.bounds}
 &&\forall j\in J^\prime:~~~\left|\lambda \sum_{i\in J} C_{ij}~x_i - \rho_j - \sum_{a=1}^m Y_{ja}~\mu_a\right| \leq L_j\\
 &&\forall j\in J^+:~~~\lambda \sum_{i\in J} C_{ij}~x_i - \rho_j - \sum_{a=1}^m Y_{ja}~\mu_a\leq -L_j\label{b1+}\\
 &&\forall j\in J^-:~~~\lambda \sum_{i\in J} C_{ij}~x_i - \rho_j - \sum_{a=1}^m Y_{ja}~\mu_a\geq L_j\label{b1-}
\end{eqnarray}
With (\ref{J1.bounds}), (\ref{Y1.bounds}), (\ref{globalmin.bounds}), (\ref{b1+}) and (\ref{b1-}), the second line in (\ref{global.min}) is positive-definite for all $\epsilon_i$ (subject to (\ref{e+}) and (\ref{e-}), that is) -- this is because these terms are linear in $\epsilon_i$. On the other hand, the first term in the first line of (\ref{global.min}) is positive semi-definite as $C_{ij}$ is assumed to be positive-definite. The second term is positive semi-definite as $\eta_i = \mbox{sign}\left(x_i\right)$ for $i\in J$. The third term implies that for any $\epsilon_a\neq 0$ we have the following condition on $\epsilon_i$:
\begin{equation}\label{const.epsilon}
 \sum_{i=1}^N \epsilon_i~Y_{ia} = 0
\end{equation}
This is simply the condition that we are only allowed to consider paths $x_i\rightarrow x_i + \epsilon_i$ along which the constraints (\ref{Y1.bounds}) are satisfied (for all $i\in\{1,\dots,N\}$).

{}The conditions (\ref{globalmin.bounds}), (\ref{b1+}) and (\ref{b1-}) must be satisfied by the solution to (\ref{J1.bounds}) and (\ref{Y1.bounds}), which give a global optimum. However, even ignoring the bounds for a moment, {\em a priori} we do not know i) what the subset $J^\prime$ is and ii) what the values of $\eta_i$ are for $i\in J$, so we have $3^N$ -- a prohibitively large number -- possible combinations.

\subsection{Optimization: Factor Model}\label{sub4.2}

{}This can be circumvented by assuming the factor model form for $C_{ij}$:
\begin{equation}
 C_{ij} = \Theta_{ij} \equiv \xi_i^2~\delta_{ij} + \sum_{A=1}^K \Omega_{iA}~{\Omega}_{jA}
\end{equation}
Here, any values of $A$ such that the corresponding column of $\Omega_{iA}$ is a linear combination of the columns of $Y_{ia}$ must be omitted (with the specific risk untouched). This is because in (\ref{J1.bounds}), (\ref{globalmin.bounds}), (\ref{b1+}) and (\ref{b1-}) $C_{ij}$ appears only in the combination
\begin{eqnarray}
 &&\sum_{j\in J} C_{ij}~x_j = \xi_i^2~x_i + \sum_{A=1}^K \Omega_{iA} \sum_{j\in J} x_j~\Omega_{jA},~~~i\in J\\
 &&\sum_{j\in J} C_{ij}~x_j = \sum_{A=1}^K \Omega_{iA} \sum_{j\in J} x_j~\Omega_{jA},~~~i\in J^\prime
\end{eqnarray}
so if any column in $\Omega_{iA}$ is a linear combination of the columns of $Y_{ia}$, its contribution vanishes due to (\ref{Y1.bounds}). We assume that all such columns in $\Omega_{iA}$, if any, are omitted.

{}The optimization problem reduces to solving a $(K+m)$-dimensional system. Let
\begin{equation}\label{v}
 v_A \equiv \sum_{i=1}^N~x_i~\Omega_{iA} = \sum_{i\in J}~x_i~\Omega_{iA},~~~A=1,\dots,K
\end{equation}
Further, let $H\equiv \{a\}\cup\{A\}$. Let ${\widehat \Omega}_{i\alpha}$, $\alpha\in H$ be the following $N\times(K+m)$ matrix
\begin{eqnarray}
 &&{\widehat \Omega}_{ia} \equiv Y_{ia}\\
 &&{\widehat \Omega}_{iA} \equiv \Omega_{iA}
\end{eqnarray}
Let $u_\alpha$ be the following $(K+M)$-vector:
\begin{eqnarray}
 &&u_a \equiv -{1\over\lambda}~\mu_a\\
 &&u_A\equiv v_A
\end{eqnarray}
{}From (\ref{J1.bounds}), (\ref{Y1.bounds}) and (\ref{v}) we have
\begin{eqnarray}\label{wv.bounds}
 &&x_i = {1\over \lambda \xi_i^2}~\left(\rho_i - L_i~\eta_i - \lambda \sum_{\alpha \in H} {\widehat \Omega}_{i\alpha}~u_\alpha\right),~~~i\in {\widetilde J}\\
 &&\sum_{i\in J}x_i~{\widehat\Omega}_{i\alpha} = \sum_{\beta\in H}\varphi_{\alpha\beta}~u_\beta \label{v1.bounds}
\end{eqnarray}
where $\varphi_{\alpha\beta}$ is the following symmetric $(K+m)\times(K+m)$ matrix:
\begin{eqnarray}
 &&\varphi_{AB} \equiv \delta_{AB}\\
 &&\varphi_{Ab} = 0\\
 &&\varphi_{ab} = 0
\end{eqnarray}
Recalling that we have
\begin{equation}\label{x-eta.bounds}
 x_i~\eta_i > 0,~~~i\in {\widetilde J}
\end{equation}
we get
\begin{eqnarray}
 &&\eta_i = \mbox{sign}\left(\rho_i - \lambda \sum_{\alpha\in H} {\widehat \Omega}_{i\alpha}~u_\alpha\right),~~~i\in {\widetilde J}\label{eta1.bounds}\\
 &&\forall i\in J^+:~~~\rho_i - \lambda \sum_{\alpha\in H} {\widehat \Omega}_{i\alpha}~u_\alpha \geq L_i + \lambda~\xi_i^2~x_i^+\equiv L_i^+\label{bounds+}\\
 &&\forall i\in J^-:~~~\rho_i - \lambda \sum_{\alpha\in H} {\widehat \Omega}_{i\alpha}~u_\alpha \leq -L_i + \lambda~\xi_i^2~x_i^-\equiv -L_i^-\label{bounds-}\\
 &&\forall i\in {\widetilde J}:~~~\left|\rho_i - \lambda \sum_{\alpha\in H} {\widehat \Omega}_{i\alpha}~u_\alpha\right| > L_i \label{eta2.bounds}\\
 &&\forall i\in J^\prime:~~~\left|\rho_i - \lambda \sum_{\alpha\in H} {\widehat \Omega}_{i\alpha}~u_\alpha\right| \leq L_i\label{eta3.bounds}
\end{eqnarray}
where (\ref{eta2.bounds}) follows from (\ref{x-eta.bounds}) and (\ref{wv.bounds}). The last four inequalities define $J^+$, $J^-$, ${\widetilde J}$ and $J^\prime$ in terms of $(K+m)$ unknowns $u_\alpha$. Note that $L_i^\pm > L_i$, $i\in J^\pm$ and if we take $x_i^\pm \rightarrow \pm\infty$, we get empty $J^\pm$.

{}Substituting (\ref{wv.bounds}) into (\ref{v1.bounds}), we get the following system of $(K+m)$ equations for $(K+m)$ unknowns $u_\alpha$:
\begin{equation}
 \sum_{\beta\in H} {\widehat Q}_{\alpha\beta}~u_\beta = y_\alpha
\end{equation}
where
\begin{eqnarray}
 &&{\widehat Q}_{\alpha\beta} \equiv \varphi_{\alpha\beta} + \sum_{i\in {\widetilde J}} {{{\widehat \Omega}_{i\alpha}~{\widehat \Omega}_{i\beta}} \over {\xi_i^2}}\\
 &&y_\alpha \equiv {1\over\lambda} \sum_{i\in {\widetilde J}} {{{\widehat \Omega}_{i\alpha}} \over {\xi_i^2}}\left[\rho_i - L_i~\eta_i\right] + \sum_{i\in J^+} x_i^+~{\widehat\Omega}_{i\alpha} + \sum_{i\in J^-} x_i^-~{\widehat\Omega}_{i\alpha}
\end{eqnarray}
so we have
\begin{equation}\label{Qv.bounds}
 u_\alpha = \sum_{\beta\in H} {\widehat Q}^{-1}_{\alpha\beta}~y_\beta
\end{equation}
where ${\widehat Q}^{-1}$ is the inverse of ${\widehat Q}$.

{}Note that (\ref{Qv.bounds}) solves for $u_\alpha$ given $\eta_i$, $J^+$, $J^-$, ${\widetilde J}$ and $J^\prime$. On the other hand, (\ref{eta1.bounds}), (\ref{bounds+}), (\ref{bounds-}), (\ref{eta2.bounds}) and (\ref{eta3.bounds}) determine $\eta_i$, $J^+$, $J^-$, ${\widetilde J}$ and $J^\prime$ in terms of $u_\alpha$. The entire system is then solved iteratively, where at the initial iteration one takes ${\widetilde J}^{(0)}=\{1,\dots,N\}$, so that $J^{+(0)}$, $J^{-(0)}$ and $J^{\prime(0)}$ are empty, and
\begin{equation}
 \eta^{(0)}_i = \pm 1,~~~i=1,\dots,N
\end{equation}
While {\em a priori} the values of $\eta^{(0)}_i$ can be arbitrary, unless $(K+m)\ll N$, in some cases one might encounter convergence speed issues. However, if one chooses
\begin{equation}
 \eta^{(0)}_i = \mbox{sign}(\rho_i),~~~i=1,\dots,N
\end{equation}
then the iterative procedure generally is expected to converge rather fast.

{}The following trick can speed up the convergence. Let ${\widehat x}_i^{(s)}$ be such that
\begin{eqnarray}
 &&\forall i\in\{1,\dots,N\}:~~~x_i^-\leq {\widehat x}^{(s)}_i\leq x_i^+\\
 &&\sum_{i=1}^N {\widehat x}^{(s)}_i~Y_{ia} = 0,~~~a=1,\dots,m
\end{eqnarray}
Let $x^{(s+1)}_i$ be the solution obtained at the $(s+1)$-th iteration. This solution satisfies the linear constraints, but may not satisfy the bounds. Let
\begin{eqnarray}
 &&q_i \equiv x^{(s+1)}_i - {\widehat x}^{(s)}_i\\
 &&h_i(t) \equiv {\widehat x}^{(s)}_i + t~q_i,~~~t\in[0,1]
\end{eqnarray}
Then
\begin{equation}
 {\widehat x}^{(s+1)}_i\equiv h_i(t_*) = {\widehat x}^{(s)}_i + t_*~q_i
\end{equation}
where $t_*$ is the maximal value of $t$ such that $h_i(t)$ satisfies the bounds. We have:
\begin{eqnarray}
 &&q_i > 0:~~~p_i \equiv \mbox{min}\left(x^{(s+1)}_i, ~x_i^+\right)\\
 &&q_i < 0:~~~p_i \equiv \mbox{max}\left(x^{(s+1)}_i, ~x_i^-\right)\\
 &&t_* = \mbox{min}\left({{p_i-{\widehat x}^{(s)}_i}\over q_i} ~\Big|~ q_i\neq 0,~i=1,\dots,N\right)
\end{eqnarray}
Now, at each step, instead of (\ref{bounds+}) and (\ref{bounds-}), we can define $J^+$ and $J^-$ via ($J^\prime$ is still defined via (\ref{eta3.bounds}))
\begin{eqnarray}
 &&\forall i\in J^+:~~~{\widehat x}_i = x_i^+\label{b+}\\
 &&\forall i\in J^-:~~~{\widehat x}_i = x_i^-\label{b-}
\end{eqnarray}
where ${\widehat x}_i$ is computed iteratively as above and we can take ${\widehat x}_i^{(0)} \equiv 0$ at the initial iteration. The difference between (\ref{b+}), (\ref{b-}) and (\ref{bounds+}), (\ref{bounds-}) is that the former add new elements to the sets $J^+$ and $J^-$ one (or a few) element(s) at each iteration, while the latter can add many elements.

{}The convergence criteria are given by (this produces the global optimum)
\begin{eqnarray}
 &&{\widetilde J}^{(s + 1)} = {\widetilde J}^{(s)}\\
 &&J^{+(s + 1)} = J^{+(s)}\\
 &&J^{-(s + 1)} = J^{-(s)}\\
 &&\forall i\in {\widetilde J}^{(s+1)}:~~~\eta^{(s+1)}_i = \eta^{(s)}_i\\
 &&\forall \alpha\in H:~~~u^{(s + 1)}_\alpha = u^{(s)}_\alpha
\end{eqnarray}
The first four of these criteria are based on discrete quantities and are unaffected by computational (machine) precision effects, while the last criterion is based on continuous quantities and in practice is understood as satisfied within computational (machine) precision or preset tolerance.

\section{Example: Intraday Mean-Reversion Alpha}\label{sec7}

{}In this section, to illustrate our discussion in Section \ref{sec.mean.rev}, we discuss an intraday mean-reversion alpha. Let us set up our notations. $P_i$, $i=1,\dots,N$ is the stock price for the stock labeled by $i$, where $N$ is the number of stocks in our universe. In actuality, the price for each stock is a time-series: $P_{is}$, $s=0,1,\dots,M$, where the index $s$ labels trading dates, with $s=0$ corresponding to the most recent date in the time series. We will use superscript $O$ (unadjusted open price), $C$ (unadjusted close price), $AO$ (open price fully adjusted for splits and dividends), and $AC$ (close price fully adjusted for splits and dividends), so, {\em e.g.}, $P^C_{is}$ is the unadjusted close price. $V_{is}$ is the unadjusted daily volume (in shares, not dollars). We define the overnight return as the close-to-next-open return:
\begin{equation}
 R_{is} \equiv \ln\left({P^{AO}_{is} / P^{AC}_{i,s+1}}\right)
\end{equation}
Note that both prices in this definition are fully adjusted.

{}Next, we take an $N\times K$ binary loadings matrix $\Lambda_{iA}$ for our universe in three incarnations, based on Bloomberg Industry Classification System (BICS) sectors, industries and sub-industries. These are binary clusters discussed in Subsection \ref{sub1.4}.\footnote{\, Note that stocks rarely jump (sub-)industries/sectors, so $\Lambda_{iA}$ can be assumed to be static.} For each date $s$, we cross-sectionally regress our returns $R_{is}$ over $\Lambda_{iA}$ with no intercept\footnote{\, More precisely, the intercept is already subsumed in $\Lambda_{iA}$: $\Lambda_{iA} = 1$ if the stock labeled by $i$ belongs to the cluster labeled by $A = 1,\dots,K$; otherwise, $\Lambda_{iA} = 0$. Each stock belongs to one and only one cluster. This implies that $\sum_{A=1}^{K} \Lambda_{iA} = 1$ for each $i$, so a linear combination of the columns of $\Lambda_{iA}$ is the intercept.} and unit weights, as in (\ref{R.Lambda}). We take the residuals $\varepsilon_{is}$ of the regression (\ref{R.Lambda}), and specify the desired dollar holdings via
\begin{eqnarray}\label{D.i.model}
 &&D_{is} = -{\varepsilon}_{is} ~ {I\over\sum_{j=1}^N \left|{\varepsilon}_{js}\right|}\\
 &&\sum_{i=1}^N \left|D_{is}\right| = I\\
 &&\sum_{i=1}^N D_{is} = 0
\end{eqnarray}
where $I$ is the {\em intraday} investment level, which is the same for all dates $s$.

{}The portfolio is established at the open\footnote{\, This is a so-called ``delay-0" alpha -- $P^O_{is}$ is used in the alpha, and as the establishing fill price.} assuming fills at the open prices $P^O_{is}$, and liquidated at the close on the same day assuming fills at the close prices $P^C_{is}$, with no transaction costs or slippage, both of which are present in real life -- here our goal is not to build a realistic trading strategy that will make money in real life, but to illustrate our discussion in Section \ref{sec.mean.rev}. Daily P\&L for each stock is given by
\begin{equation}
 \Pi_{is} = D_{is}\left[{P^C_{is}\over P^O_{is}}-1\right]
\end{equation}
The shares bought plus sold ({\em i.e.}, for the establishing and liquidating trades combined) for each stock on each day are computed via $Q_{is} = 2 |D_{is}| / P^O_{is}$.

{}{}Before we can run our regressions, we need to select our universe. We wish to keep our discussion here as simple as possible, so we select our universe based on the average daily dollar volume (ADDV) defined via
\begin{equation}\label{ADDV}
 A_{is}\equiv {1\over d} \sum_{r=1}^d V_{i, s+r}~P^C_{i, s+r}
\end{equation}
We take $d=21$ ({\em i.e.}, one month), and then take our universe to be top 2000 tickers by ADDV. However, to ensure that we do not inadvertently introduce a universe selection bias,\footnote{\, {\em I.e.}, to ensure that our results are not a mere consequence of the universe selection.} we do not rebalance the universe daily. Instead, we rebalance monthly, every 21 trading days, to be precise. {\em I.e.}, we break our 5-year backtest period (see below) into 21-day intervals, we compute the universe using ADDV (which, in turn, is computed based on the 21-day period immediately preceding such interval), and use this universe during the entire such interval.\footnote{\, Note that, since the alpha is purely intraday, this ``rebalancing" does not generate additional trades, it simply changes the universe that is traded for the next 21 days.} The bias that we do have, however, is the survivorship bias. We take the data for the universe of tickers as of 9/6/2014 that have historical pricing data on http://finance.yahoo.com (accessed on 9/6/2014) for the period 8/1/2008 through 9/5/2014. We restrict this universe to include only U.S. listed common stocks and class shares (no OTCs, preferred shares, {\em etc.}) with BICS sector, industry and sub-industry assignments as of 9/6/2014.\footnote{\, The number of such tickers in our data is 3,811. The number of BICS sectors is 10. The numbers of BICS industries is 48.  The number of BICS sub-industries varies between 164 and 169 (due to small sub-industries, which are affected by the varying top-2000-by-ADDV universe).} However, it does not appear that the survivorship bias is a leading effect here (see below). Also, ADDV-based universe selection is by no means optimal and is chosen here for the sake of simplicity. In practical applications, the trading universe of liquid stocks is carefully selected based on market cap, liquidity (ADDV), price and other (proprietary) criteria.

{}We run our simulation over a period of 5 years. More precisely, $M = 252\times 5$, and $s=0$ is 9/5/2014 (see above). The results for the annualized return-on-capital (ROC), annualized Sharpe ratio (SR) and cents-per-share (CPS) are given in Table \ref{table1} for 3 choices of clusters: BICS sectors, industries and sub-industries. ROC is computed as average daily P\&L divided by the investment level $I$ (with no leverage) and multiplied by 252. SR is computed as daily Sharpe ratio multiplied by $\sqrt{252}$. CPS is computed as the total P\&L divided by total shares traded. The P\&L graphs for the 3 cases in Table \ref{table1} are given in Figure 1.

{}In the above model we have done no risk management apart from (automatic in this case) dollar neutrality. We can do risk management via weighted regression as in Subsection \ref{sub1.7}. However, here we will discuss another method. The basic issue is that some residuals $\epsilon_{is}$ can be very large, so the strategy can disproportionately load up on stocks with such large residuals and as a result the portfolio is not diversified enough. This is why SR in Table \ref{table1} are not as high as in Table \ref{table2} (see below). We can deal with such large residuals by treating them as outliers. One well-known method is Winsorization. Here we discuss a conceptually similar method, which is more convenient. Let $X_i$ be a set of values for which we expect to have a normal distribution with cross-sectional mean ${\overline X}$ and standard deviation $\chi$. Let ${\widetilde X}_i$ be the values of $X_i$ deformed such that ${\widetilde X}_i$ are conformed to the normal distribution with the same mean ${\overline X}$ and standard deviation $\chi$. {\em E.g.}, we can use the {\tt normalize()} function given in Appendix A of (Kakushadze and Liew, 2014). Now let us apply this method to our residuals $\varepsilon_{is}$ separately for each date (so everything is out-of-sample). Let the resulting values be ${\widetilde \varepsilon}_{is}$. Note that ${\widetilde \varepsilon}_{is}$ still have vanishing cross-sectional means, but the outliers have been ``squashed". We can now use ${\widetilde \varepsilon}_{is}$ instead of $\varepsilon_{is}$ in (\ref{D.i.model}) and maintain dollar neutrality. The results for ROC, SR and CPS are given in Table \ref{table2}. Note a dramatic increase in SR compared with Table \ref{table1} -- at the expense of lowering ROC and CPS. The P\&L graphs for the 3 cases in Table \ref{table2} are given in Figure 2.

{}One evident caveat of this alpha is that the ``open" can be a fuzzy notion as some stocks do not always open at 9:30:00 sharp. So, our assumption that the orders can be placed simultaneously at the open is a bit faulty.\footnote{\, Plus we are assuming ``delay-0", meaning, we can place the trades ``infinitely" fast right after receiving the opening prints from the exchange(s) and get filled at the very same open prices. And, as mentioned above, we are ignoring trading costs and slippage, hence the rosy ROC, SR and CPS.} In real life one would have to wait until a little after the open and compute the alpha for the stocks that are open as of that time, then send the orders and get fills. An intraday simulated strategy of this type can be accessed freely on http://vynance.com/portfolio.html. The establishing time is 9:31:30 (and the liquidating time is 15:59:00). The trading universe varies from day-to-day and is smaller than 2000 tickers, it is roughly in the range of 200-300 tickers for long positions and 200-300 tickers for short positions. The performance for this strategy from 2/18/2014 through 9/19/2014 is as follows: ROC = 29.19\%; SR = 12.13; CPS = 1.52. The simulation assumes no trading costs or slippage; however, it is not a ``delay-0" but (more realistic) ``delay-30-seconds" strategy, {\em i.e.}, the alpha used for the establishing trades at 9:31:30 is computed based on the pricing data from 9:31:00 (which data itself is delayed 5-35 seconds). Since ``inception" (4/14/2011) Vynance Portfolio has had a consistent simulated performance with the annualized daily Sharpe ratio of about 16 and monthly return-on-capital of about 3\%. This is consistent with our results in Table \ref{table2}, which indicates that the survivorship bias in our results should not be a leading effect -- Vynance Portfolio simulations are done daily, in real time, and thus are free from the survivorship bias.

\section{Concluding Remarks}

{}As mentioned earlier, there are a myriad ways of doing mean-reversion. In the quantitative framework we discussed in these notes, a mean-reversion model is essentially defined by the risk factors used as loadings in regressions along with regression weights, or, in optimization, by the choice of the multi-factor risk model and constraints -- the latter usually also relating to neutrality w.r.t. some risk factors.

{}Here we should emphasize that there are all kinds of bells and whistles one can add to tweak a particular mean-reversion model, even within the aforementioned framework. Also, if only regressions are used, then factor covariance matrix is not needed and one can settle for using, {\em e.g.}, historical volatilities in regression weights, {\em i.e.}, in this case one only needs the unrotated factor loadings matrix ${\widetilde\Omega}_{iA}$. On the other hand, for optimization a full multi-factor risk model is required -- not just factor loadings matrix, but also the factor covariance matrix and specific risk. Since depending on the choice of the constraints -- and also the returns used -- one may omit some risk factors from the factor loadings matrix, in many cases it is warranted to compute a custom multi-factor risk model. This topic is covered in much more detail in (Kakushadze and Liew, 2014) dedicated to this subject.

%\newpage
\begin{table}[ht]
\caption{Simulation results for the intraday mean-reversion alphas discussed in Section \ref{sec7} without normalizing the regression residuals.} % title of Table
%\centering % used for centering table
\begin{tabular}{l l l l} % centered columns (4 columns)
\\
\hline\hline %inserts double horizontal lines
Clusters & ROC & SR & CPS\\[0.5ex] % inserts table
%heading
\hline % inserts single horizontal line
BICS Sectors & 44.58\% & 6.21 & 1.17\\
BICS Industries & 49.00\% & 7.15 & 1.29\\
BICS Sub-industries& 51.77\% & 7.87 & 1.36\\ [1ex] % [1ex] adds vertical space
\hline %inserts single line
\end{tabular}
\label{table1} % is used to refer this table in the text
\end{table}

\begin{table}[ht]
\caption{Simulation results for the intraday mean-reversion alphas discussed in Section \ref{sec7} with normalizing the regression residuals.} % title of Table
%\centering % used for centering table
\begin{tabular}{l l l l} % centered columns (4 columns)
\\
\hline\hline %inserts double horizontal lines
Clusters & ROC & SR & CPS\\[0.5ex] % inserts table
%heading
\hline % inserts single horizontal line
BICS Sectors & 33.27\% & 11.55 & 1.02\\
BICS Industries & 37.67\% & 15.02 & 1.15\\
BICS Sub-industries& 40.40\% & 18.50 & 1.24\\ [1ex] % [1ex] adds vertical space

\hline %inserts single line
\end{tabular}
\label{table2} % is used to refer this table in the text
\end{table}

\begin{figure}[ht]
\centerline{\epsfxsize 4.truein \epsfysize 4.truein\epsfbox{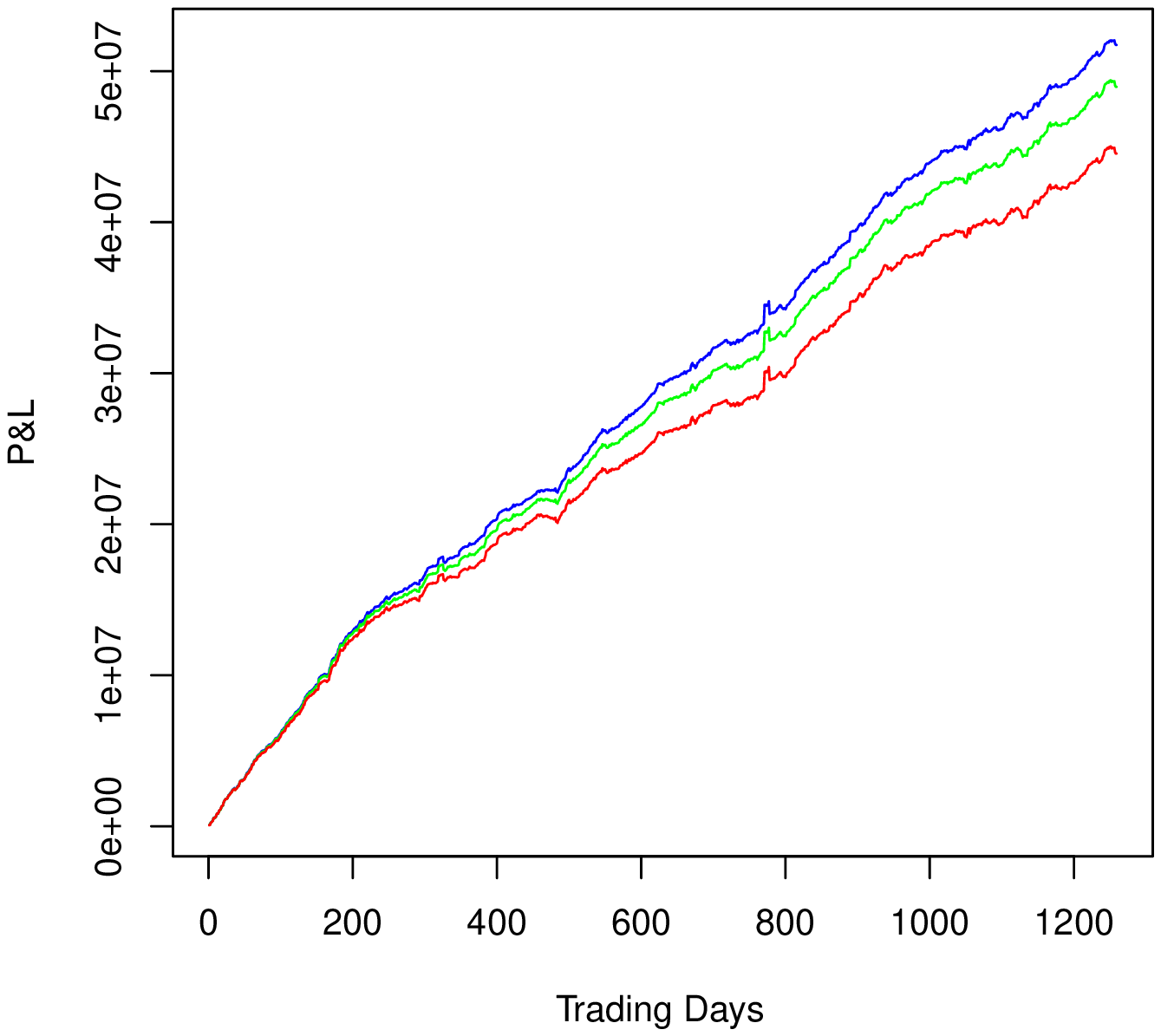}}
\noindent{\small {Figure 1. P\&L graphs for the mean-reversion alpha (unnormalized residuals) discussed in Section \ref{sec7}, with a summary in Table \ref{table1}. Bottom-to-top-performing: i) BICS sectors, ii) BICS industries, and iii) BICS sub-industries. The investment level is \$10M long plus \$10M short.}}

\end{figure}

\begin{figure}[ht]
\centerline{\epsfxsize 4.truein \epsfysize 4.truein\epsfbox{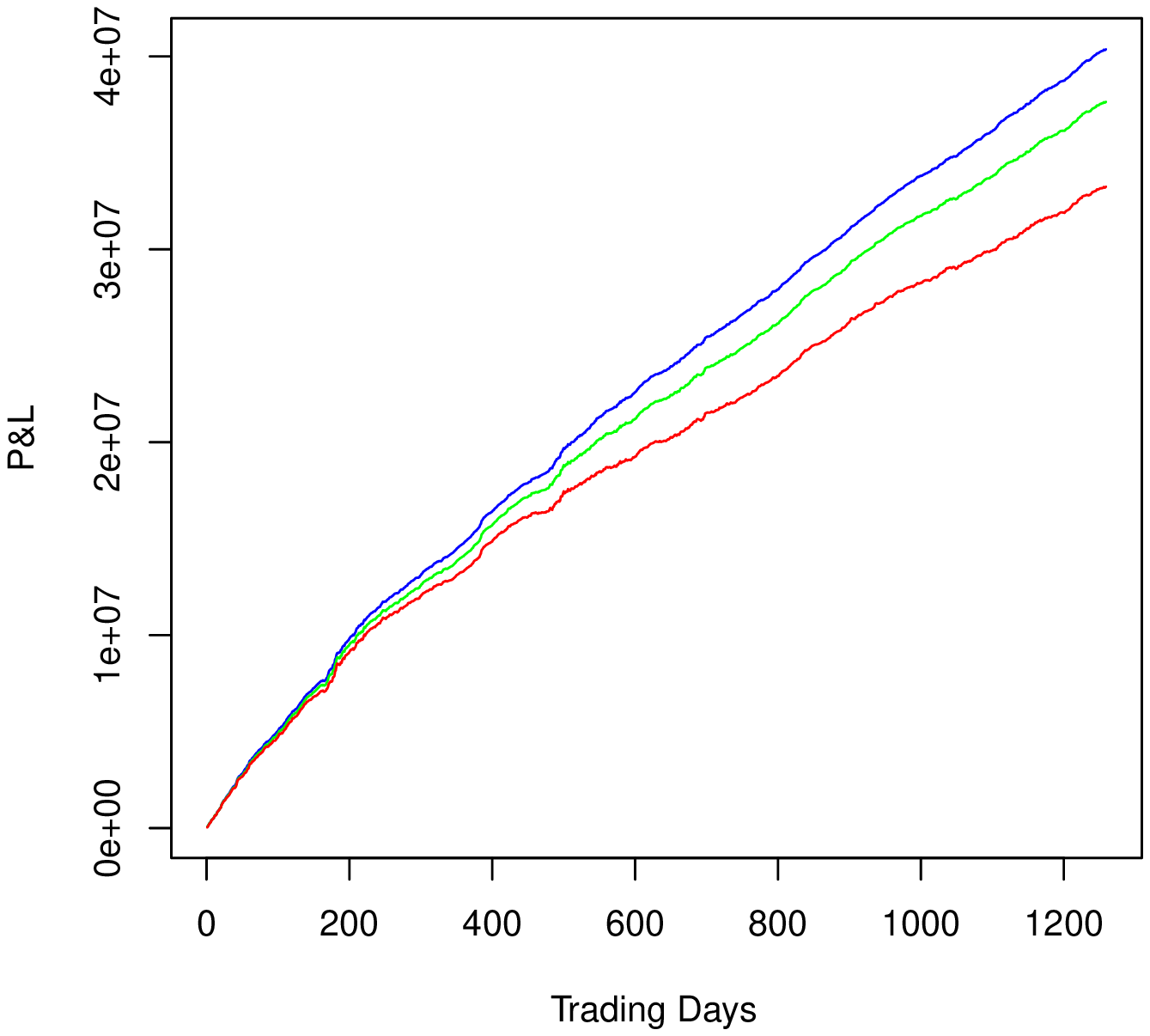}}
\noindent{\small {Figure 2. P\&L graphs for the mean-reversion alpha (normalized residuals) discussed in Section \ref{sec7}, with a summary in Table \ref{table2}. Bottom-to-top-performing: i) BICS sectors, ii) BICS industries, and iii) BICS sub-industries. The investment level is \$10M long plus \$10M short.}}

\end{figure}

\end{document}